\documentclass[twocolumn,showpacs,preprintnumbers,aps,amsmath,amssymb]{revtex4}

\usepackage{epsf}
\usepackage{rotating}

\usepackage{graphicx,epsfig}
\usepackage{dcolumn}
\usepackage{bm}

\begin{document}

\preprint{}

\title[Title]
{Models of Financial Markets with Extensive Participation Incentives}

\author{C. H. Yeung$^{1}$, K.~Y.~Michael Wong$^{1}$,
and Y.-C.~Zhang$^{1,2}$}
\affiliation{$^{1}$ Department of Physics,
The Hong Kong University of Science and Technology, Hong Kong, China\\
$^{2}$ D\'epartement de Physique,
Universit\'e de Fribourg, P\'erolles, Fribourg, CH-1700 Switzerland
}

\date{\today}

\begin{abstract}
We consider models of financial markets
in which all parties involved find incentives to participate.
Strategies are evaluated directly by their virtual wealths.
By tuning the price sensitivity and market impact,
a phase diagram with several attractor behaviors
resembling those of real markets emerge,
reflecting the roles played by the arbitrageurs and trendsetters,
and including a phase with irregular price trends and positive sums.
The positive-sumness of the players' wealths
provides participation incentives for them.
Evolution and the bid-ask spread provide mechanisms
for the gain in wealth of both the players and market-makers.
New players survive in the market if the evolutionary rate is sufficiently slow.
We test the applicability of the model on real Hang Seng Index data
over 20 years.
Comparisons with other models show that our model has a superior
average performance when applied to real financial data.

\end{abstract}

\maketitle


\section{Introduction}
\label{secIntroduction}

The potentials of using agent-based models to analyze and elucidate the
behavior of financial markets are gradually realized in the physics community
in recent years~\cite{challet_book2005}.
These models are able to relate the attributes of the indvidual players,
such as their memory sizes and payoffs,
to the collective behavior of the system,
such as the information content, 
volatility, 
and phase transitions
\cite{challet1997, savit1999, jefferies2001, andersen2003}.
The key insight of a family of these models, 
known as the Minority Games,
is the observation that the players making the minority decisions take
advantages of their counterparts during trading,
and the adaptive behavior towards
this target results in the non-trivial behavior of the financial markets.

However, 
the original Minority Game does not provide a correct perspective to 
model financial markets,
in which players participate only when they see chances of getting profits.
Since the winners in the games belong to the minority group
at each time step,
they become negative-sum games.
Consequently,
were the players given the option to withdraw from the market,
their participation cannot sustain.

Variations of the original version of the game
provided partial solutions to the issue.
For example,
the influence of wealth in the decisions of the agents was considered 
\cite{giardina2003, challet2001b, slanina1999}.
In these models, 
the buying power of an agent is curtailed when she does not have enough wealth.
This creates a natural mechanism to expel the poor agents from the market,
although the issue of participation incentives was not the focus of those studies.

Another mechanism for the agents to change their status of participation 
was considered in the so-called grand-canonical Minority Game
\cite{giardina2003, challet2001b, slanina1999, challet2000b}.
The speculators have the option to participate or withdraw from the market
according to their perception of profitability.
However, the wealth gained by the so-called {\it producers},
whose predictable strategies made them the prey of arbitrage
by the speculators, is at best zero.
Thus, the issue merely shifts from the speculators to the producers:
were the producers given the option to withdraw from the market,
their participation cannot sustain.

The $\$$-game considered whether the market makers can be prevented
from being arbitraged by the players~\cite{andersen2003}.
It was found that the market-makers are able to arbitrage the players
when the strategies of the players are too complex,
but they are arbitraged by the players otherwise.
The issue of voluntary participation remains:
either the market-makers or the players are tempted
to withdraw from the game,
were they given the option.

The agent-based models are remote from realistic financial markets
in another aspect.
More often, agents in real markets evaluate their strategies
in real financial terms, namely,
how much wealth the strategies would have brought them
in the market history,
had the players adopted them.
In contrast,  players in the agent-based models
use various ways to update the {\it virtual points}
or {\it scores} of their strategies when they make transactions.
Typical virtual point updating rules,
such as those in the original Minority Game \cite{challet1997},
evaluate the buying and selling decisions at a time step,
regardless of the need to update
the historical effects of the previous decisions.
In other models, one-step expectations of the players are considered,
leading to the $\$$-Game \cite{andersen2003},
and markets with a mixture of majority and minority game players
\cite{marsili2001}.
However, these models of myopic players
may not reflect the history-dependent considerations of real market players.

An example is the evaluation of {\it holding} a position
(that is, a decision of taking no buying or selling actions).
In real markets, players need the option to hold a position,
either by actively including it
as a decision prescribed by their strategies,
or by passively refraining from buying and selling decisions,
consequential to their reaching maximum or minimum positions respectively.
Virtual point updates in most agent-based models 
are neutral to holding positions.

To incorporate the option of holding a position, 
the most common method used in the grand-canonical models
is to monitor whether the virtual points
passively fall below a certain threshold~\cite{challet2000b, jefferies2001}. 
Now if holding a position is part of a player's option, 
the benefits of taking such a position
should be reflected in the payoff functions,
so that the players can choose to hold long (short) positions when 
prices are rising (falling). 
These considerations are best reflected in monitoring the wealth
associated with the strategy decisions.
We therefore focus on wealth-based strategies in this paper.
Wealth-based strategies were previously considered
in exogenous markets \cite{challet2005}.
As will be shown, agents with wealth-based strategies in endogenous markets
can self-organize to exhibit even more interesting coordinated behaviors.

In this paper, we consider models of financial markets
in which all parties involved find incentives to participate. 
Our models incorporate several realistic features recently
added to the Minority Game,
including wealth-based strategies and the 
opening and closing of an agent's position during trading \cite{challet2005, ren2006},
but the focus will be the issue of participation incentives,
attractor behavior,
and tests with real data. 
As will be shown,
the players are able to arbitrage the market-makers
in the absence of the spread in the bid and ask prices,
provided that the market impact
or the price sensitivity to the excess demand are sufficiently low.
On the other hand, in an evolving market,
underperforming players leave the market
and new players bring their wealth into the market,
rendering it a positive-sum game.
The market-makers share the wealth
by imposing spreads in the bid and ask prices.
To complete the picture,
we find that it is possible for the new players
to have incentives to participate,
if the evolutionary rate is sufficiently slow.
Compared with previous models,
our models are relatively simple in terms of the number of parameters,
and may be eventually amenable to analytical approaches 
similar to those which have proved successful in the 
study of the Minority Game 
\cite{marsili2000, challet2000a, heimel2001}.

The paper is organized as follows.
After introducing the model in Section \ref{secModel},
we describe in Section \ref{secPositiveSum}
the different attractor behaviors resembling those in real markets 
and map the positive-sumness of the players
in the absence evolution and the bid-ask spread.
In Section \ref{secTC} we consider the effects
of the bid-ask spread imposed by the market-makers.
In Section \ref{secEvolution}, 
we complete the picture by considering the conditions
under which the new players in an evolving market
are able to enhance their survival.
In Section \ref{secHSI}, 
we test the applicability of wealth-based strategies
on real Hang Seng index data,
followed by a conclusion in Section \ref{secConclusion}.

\section{The Model}
\label{secModel}

We consider a model of the financial market in which $N$ agents trade.
Although we are interested in financial markets in general 
(stocks, foreign exchange etc.),  
we will use the language of stock markets for the sake of convenience.
At each time step the agents make trading decisions $+1, -1$ or 0, 
which represent buying or selling a unit of stock, 
or taking a holding position respectively.
Denoting the decision of agent $i$ at time $t$ by $a_i(t)$, 
the {\it position} $k_i(t)$ of agent $i$ at time $t$ is given by 
\begin{eqnarray}
\label{position}
        k_i(t) = \sum_{t'\le t}a_i(t').
\end{eqnarray}
Positive and negative $k_i(t)$
refer to a {\it long} and {\it short} position respectively.

Introducing the positions of the agents provides a simple way to model the effects of 
the limited weatlh of the agents in real markets.
In previous models,
the amount of stocks bought (sold) by an agent following each decision
is a fraction of her capital (stock) \cite{giardina2003, challet2001b}.
Here,
to decorporate the limitations of wealth in the presence of both long
and short positions, 
we define a maximum position $K$ by imposing the constraint, 
i.e. $|k_i(t)| \leq K$.
Once the maximum position is reached, 
decisions which further increase the magnitude of the positions are ignored.
A consequence of this model of trading behavior is that wealth
accumulation is additive,
in contrast to other models in which wealth accumulation is multiplicative
\cite{giardina2003, challet2001b}.

The stock price should evolve according to their demand and supply.
Thus, the price of one unit of stock is updated by 
\begin{eqnarray}
\label{price}
	P(t+1) = P(t) + {\rm sign}[A(t)]|A(t)|^\gamma
\end{eqnarray}
where $A(t)\equiv \sum_i a_i(t)$ represent the excess demand.
When there are more (less) buyers than sellers,
$A(t)$ is positive (negative) driving the price up (down)
according to Eq.~(\ref{price}).
The exponent $\gamma$ describes
the {\it sensitivity} of price increment to the excess demand,
$\gamma=0$ and 1 for the cases of step
\cite{challet1997, savit1999, jefferies2001} 
and linear functions \cite{challet2000b} respectively,
which was extensively studied in the literature.
On the other hand, 
there is evidence for $\gamma \approx 0.5$ \cite{sqrt}.

In market clearing processes,
there is usually a discrepancy
between the price expected by an agent when she submitted her bid
and the actual transaction price \cite{challet2005}.
This discrepancy arises from the sequential process
of clearing the deals at each time step, 
during which the transaction price changes
from its present value to the new value according to Eq.~(\ref{price}).
For simplicity, we assume that the transacion price
is the same for all transactions at time $t$.
Thus, we approximate the transaction price $P_{\rm T}$ as 
\begin{eqnarray}
\label{transactionPrice}
	P_{\rm T}(t) = (1-\beta)P(t) + \beta P(t+1)
\end{eqnarray}
with $0 \leq \beta \leq 1$.
The variable $\beta$ acts as a {\it market impact} factor, 
since it arises from the collective effects of agent participation, 
and reduces the profit of the majority of the participating agents
relative to the so-called price takers. 
When $\beta=1$, 
agents experience {\it full} market impact in transactions 
and are trading with the {\it next price}.
When $\beta=0$,
agents experience {\it no} market impact in transactions 
and are trading with the {\it immediate price}.

With this evolving transaction price, 
we let $w_i(t)$ be the 
wealth of agent $i$ at time $t$ 
just after the decision $a_i(t)$ is carried out
and the transaction is completed.
The wealth $w_i(t)$ is the sum of cash $c_i(t)$
and stock values in hand, namely, 
\begin{eqnarray}
\label{wealth}
	w_i(t) = c_i(t) + k_i(t)P_{\rm T}(t), 
\end{eqnarray}
with $k_i(t)$ given by Eq.~(\ref{position}).
The second term of  Eq.~(\ref{wealth})
corresponds to the stock values in hand,
which is the product of the position held
and the current transaction price.
$P_{\rm T}(t)$ instead of $P(t)$ is employed in calculating wealth,
since it represents the actual stock value once decisions are implemented. 
Cash is updated according to the buying and selling of stocks at time $t$, 
namely,
\begin{eqnarray}
	c_i(t) = c_i(t-1) - a_i(t)P_{\rm T}(t).
\end{eqnarray}
Suppose agent $i$ buys (sells) a unit of stock at time $t$.
Her cash is lowered (raised) by an amount $P_{\rm T}(t)$
while her value of stocks in hand is increased by the same amount,
so that any change in wealth is due to
the change in value of the stocks she previously held.
After rearranging Eqs.(\ref{price})-(\ref{wealth}), 
the wealth change of agent $i$ after transactions at time $t$
can be expressed as
\begin{eqnarray}
\label{wealthChange}
	w_i(t)-w_i(t-1) = k_i(t-1)[P_{\rm T}(t)-P_{\rm T}(t-1)].
\end{eqnarray}

Next, we consider the strategies used by the agents
to reach their decisions in the evolving market environment. 
In this paper we will consider {\it endogenous} environments.
In this case, the market environmment at time $t$ is described by $\mu(t)$,
which is one of the $2^m$ strings
of the $m$ most recent outcomes of the sign of price changes.
Thus, $m$ is called the {\it memory size}.

A strategy prescribes the decisions $+1, 0, -1$
in response to each of the $2^m$ states of the market environment. 
Each agent draws $s$ strategies randomly at the beginning of the game.
Every strategy should have at least one buying and selling decisions.
At every time step,
each agent selects the most successful strategy
among the $s$ strategies she owns
and uses it to make a decision.
The success of a strategy is measured
by the virtual wealth it should have acquired
were its decisions followed in the market history.
This means that for strategy $\sigma$, 
we start with the initial virtual wealth $w_\sigma(0) = 0$
and update their values in the same way as we do for real wealth, that is, 
\begin{eqnarray}
\label{strategyWealth}
	k_\sigma(t) = \max\{-K, \min[K, k_\sigma(t-1)+a_\sigma(t)]\},
	\\ 
	w_\sigma(t) = w_\sigma(t-1)+k_\sigma(t-1)[P_{\rm T}-P_{\rm T}(t-1)], 
\end{eqnarray}
where $P_{\rm T}$ is the transaction price determined from 
the real transactions in Eq.~(\ref{transactionPrice}).
In this way, the virtual wealths of the strategies are calculated
as if they were the price takers.

A common way to model the incentives of the agents to participate in the market
or otherwise is to allow one of their $s$ strategies to be a 0-strategy,
such as that adopted in the grand-canonical Minority Game
\cite{giardina2003, challet2001b, slanina1999, challet2000b}. 
The 0-strategy refers to one with a holding decision irrespective 
of the market environment.
Indeed, 
this feature can be incorporated into our model.
However, 
as we shall see,
the basic behavior remains the same.
Hence,
unless mentioned explicitly,
the 0-strategy will not be included in our model.

Unlike the original Minority Game~\cite{challet1997}, 
an agent in this model can hold a long or short position
while at the same time the virtual positions of the adopted strategy
are updated independently.
The use of virtual wealths in evaluating strategies
is an essential difference with the conventional models of the Minority Game, 
which will be discussed in detail in Section~\ref{secHSI}.  
The ways of wealth accounting in the current market model 
and the Minority Game are summarized in Table~\ref{tableWealth}
in Section~\ref{secHSI}.
In the present model, the increment of real wealth
as well as the virtual wealths of strategies at each step
depends on the historical sum of actions
according to Eqs.(\ref{position}), (\ref{wealthChange}),
and Eqs.(\ref{strategyWealth}), (8) respectively,
which implicitly embed a longer memory scale in wealth monitoring
and strategy selection than the original Minority Game.
This long memory in wealh monitoring
introduces a great complexity in the dynamics of system,
which brings in a rich behavior reflecting aspects of real markets.

\section{The Phase Diagram}
\label{secPositiveSum}

\begin{figure}
\centerline{\epsfig{figure=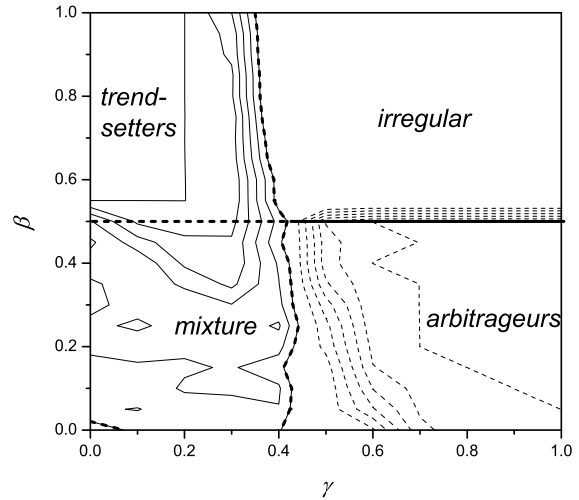, width=\linewidth}}
\caption{
The phase diagram showing the phases of the trendsetters'
attractor, the arbitageurs' attractor, 
the mixture phase, and the irregular phase for 
$m=3$, $s=2$, $K=3$, $N=1000$, $10^6$ steps and 100 samples.
Thick solid (dashed) line: transition discontinuous (continuous)
in predictability and volatility.
Thin solid (dashed) line: contours of the probability
of the trendsetters' (arbitrageurs') attractor.
}
\label{attractorPhaseDia}
\end{figure}

\begin{figure*}
\centering
\leftline{\epsfig{figure=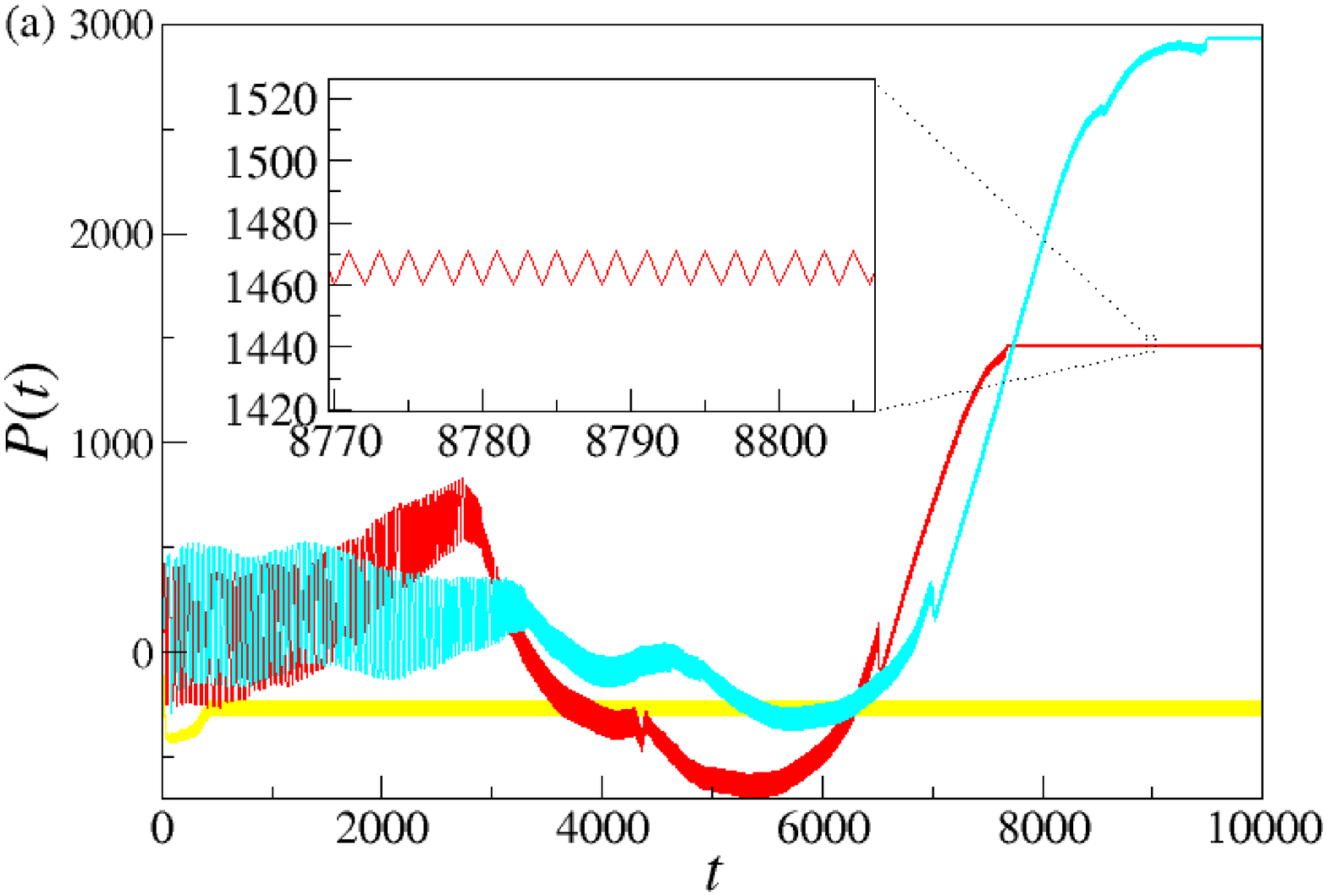, width=0.47\linewidth}                                                        
\leftline{\epsfig{figure=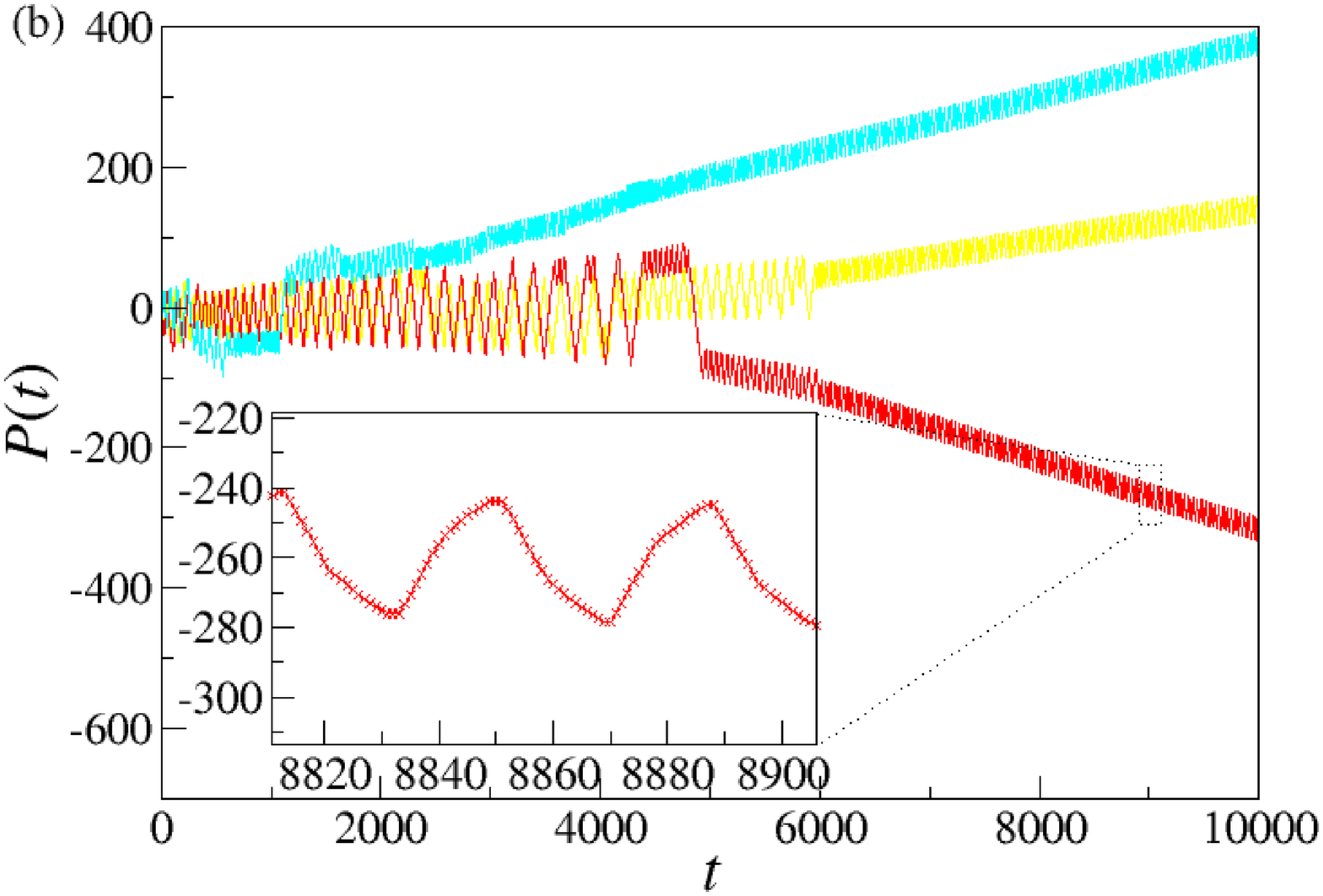, width=0.47\linewidth}}}
\leftline{\epsfig{figure=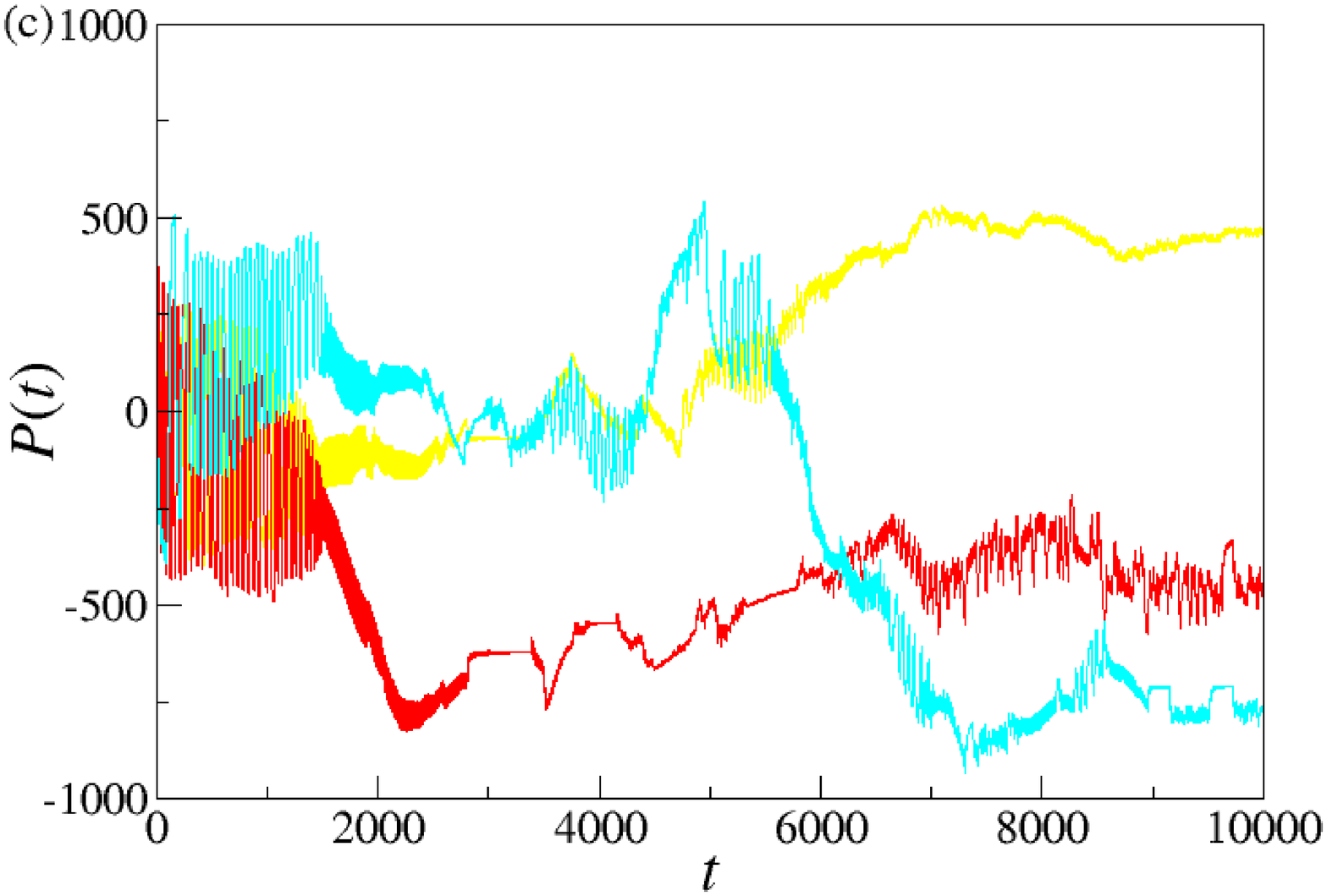, width=0.47\linewidth}                                                        
\leftline{\epsfig{figure=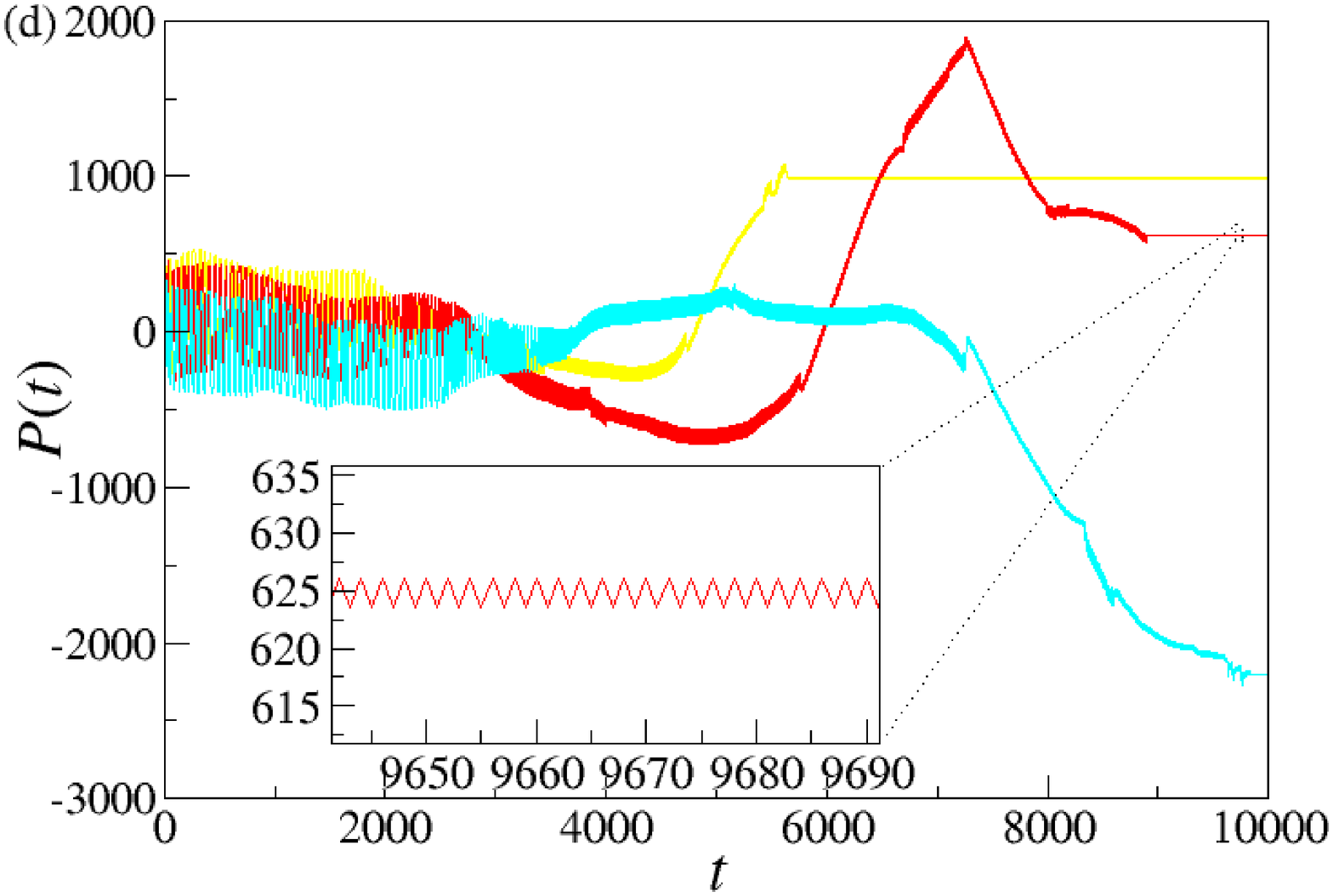, width=0.47\linewidth}}}
\caption{
(Color online) The time dependence of price for 3 samples with 
$m=3$, $s=2$, $K=1$, $N=1000$ for
(a) the arbitrageurs' phase at $(\gamma, \beta) = (0.8, 0.4)$,
(b) the trendsetters' phase at $(\gamma, \beta) = (0.2, 0.8)$,
(c) the irregular phase at $(\gamma, \beta) = (0.8, 0.8)$, and
(d) the phase boundary at $(\gamma, \beta) = (0.8, 0.5)$.
}
\label{allPrice}
\end{figure*}

In the phase space of price sensitivity $\gamma$ and market impact $\beta$,
we found the phases shown in Fig.~\ref{attractorPhaseDia}.
Examples of the time series of price and the corresponding average wealth
are shown respectively in Figs.~\ref{allPrice} and 3.
It can be seen that the system behaves very differently
in the arbitrageurs', the trendsetters' and the irregular phases.
We will discuss these behaviors in detail in the following subsections.

\begin{figure*}
\centering
\leftline{\epsfig{figure=w0804.eps, width=0.47\linewidth}                                                        
\leftline{\epsfig{figure=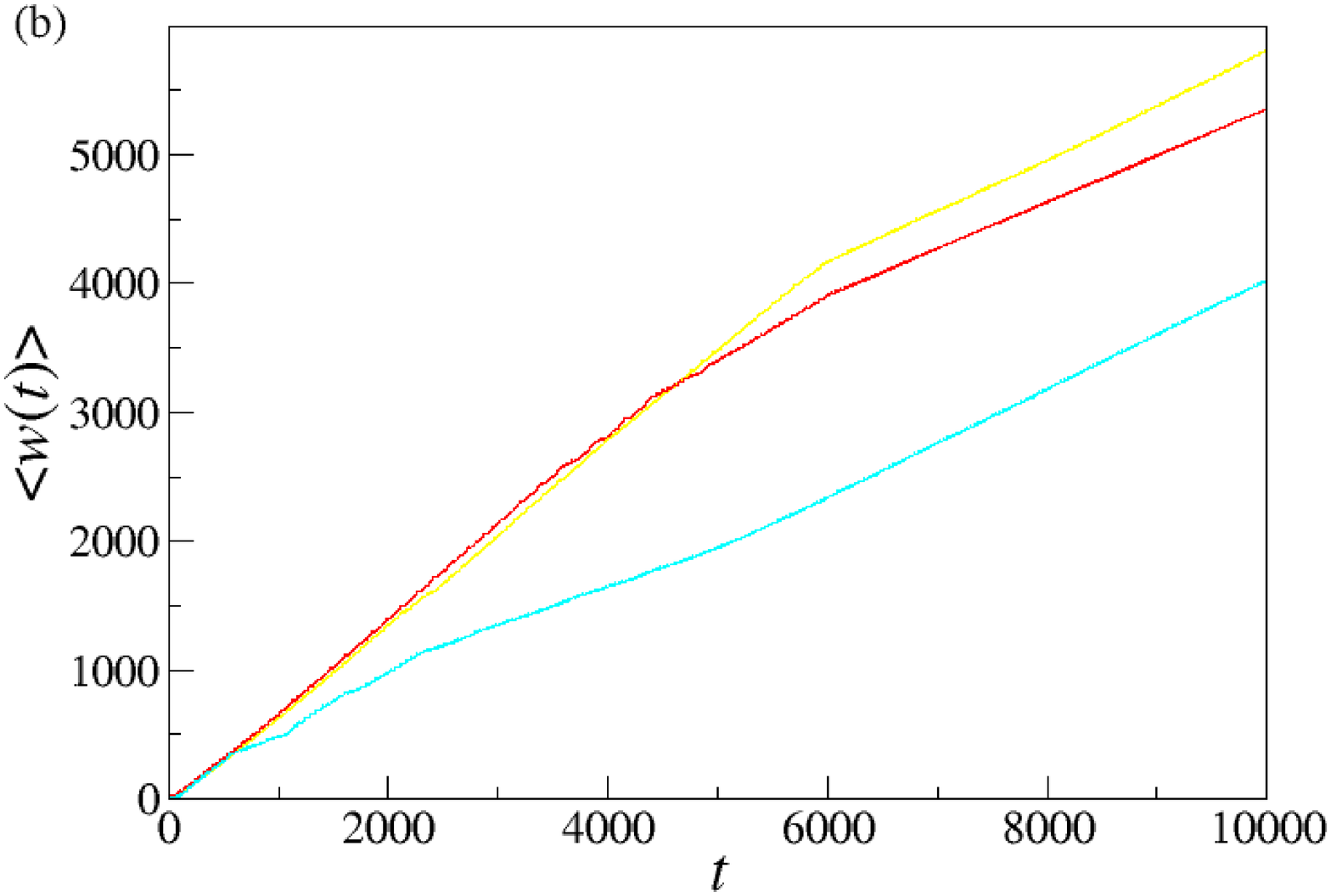, width=0.47\linewidth}}}
\leftline{\epsfig{figure=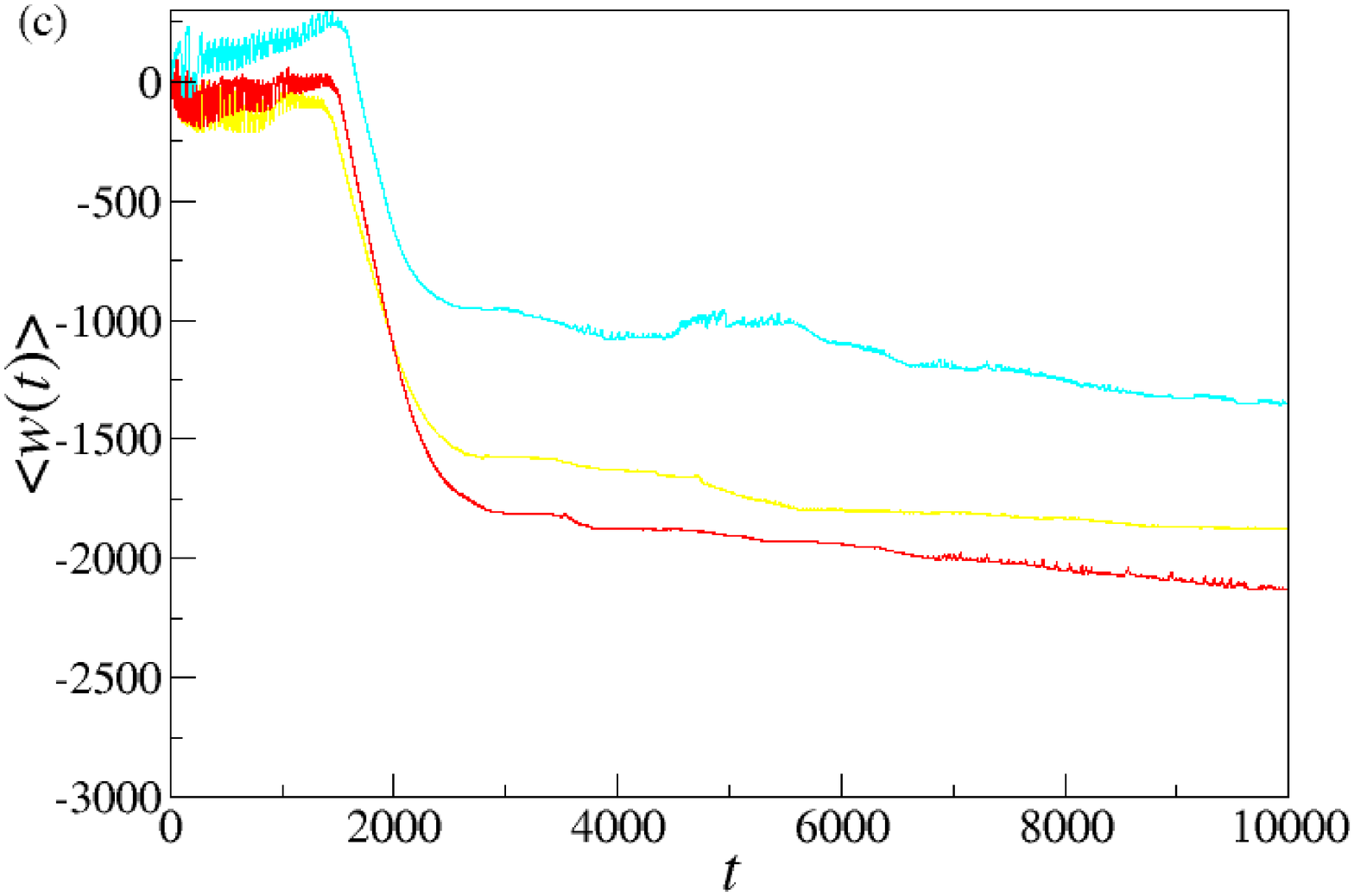, width=0.47\linewidth}                                                        
\leftline{\epsfig{figure=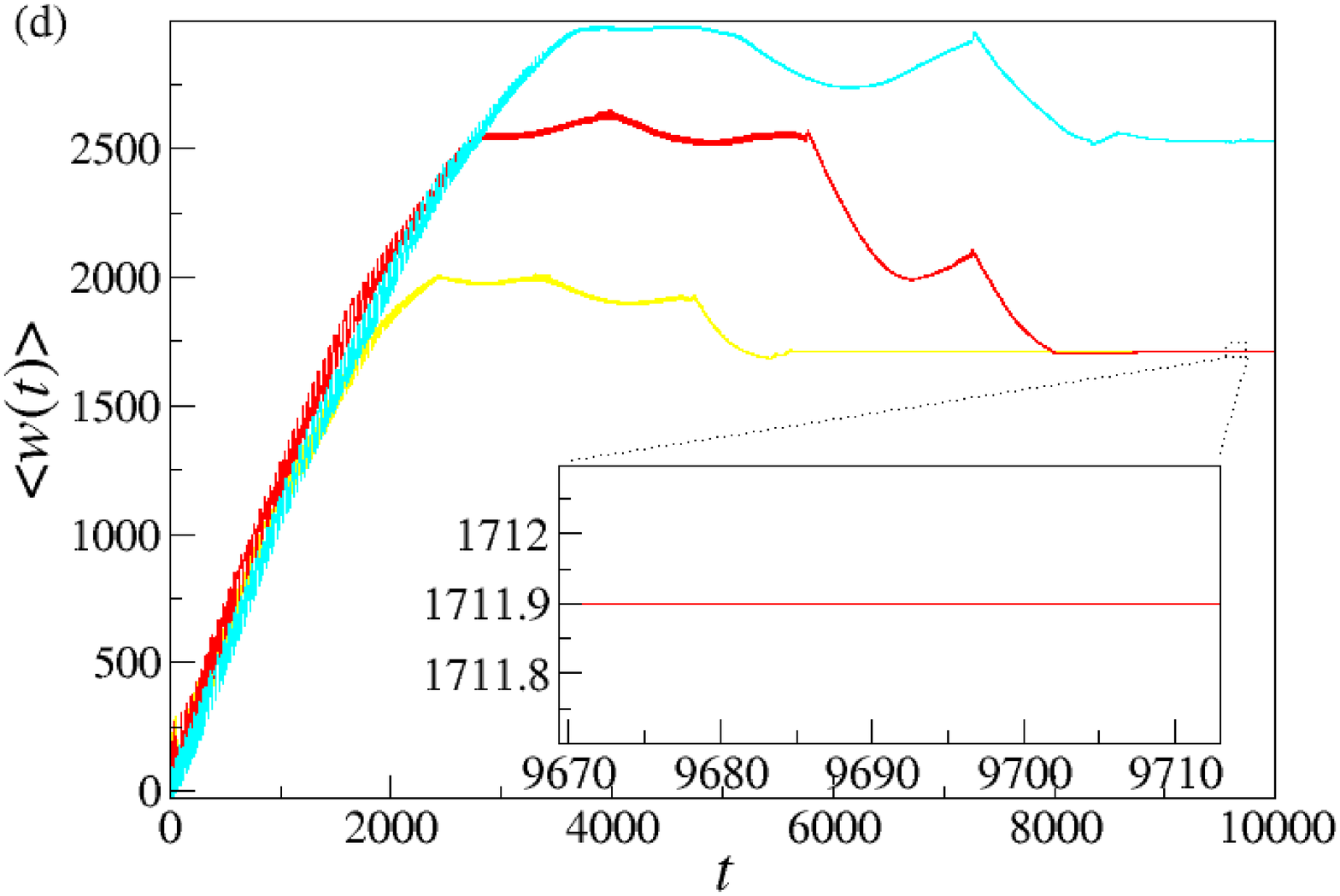, width=0.47\linewidth}}}
\caption{
(Color online) The time dependence of the average wealth for the corresponding samples
in Fig.~\ref{allPrice}.
}
\label{allWealth}
\end{figure*}

\subsection{The Arbitrageurs' Attractor}

In the region with $\beta \leq 0.5$ and sufficiently large $\gamma$, 
the system eventually falls into periodic attractors
with agents buying and selling
according to the self-generated price change in a recursive manner.
In most cases, the system gets attracted to a period-two cycle
in which the same groups of agents synchronize
into buying and selling at consective time steps,
which drives the price oscillating
between a higher and lower value accordingly,
as shown in Fig.~\ref{allPrice}(a).
Suppose the prices in the period-2 cycle are $P_0$ and $P_0+|A_0|^\gamma$.
Then the transaction prices become $P_0+\beta|A_0|^\gamma$ and 
$P_0+(1-\beta)|A_0|^\gamma$.
With $\beta<0.5$,
agents engaged in this cycle gain $(1-2\beta)|A_0|^\gamma$ per cycle by 
buying at a lower price and selling immediately at a higher price.
These agents can be called the {\it arbitrageurs}.
The attractor is robust since the net gain in wealth of the arbitrageurs
reinforces their use of the so-called {\it smart strategies}
when the market dynamics is dominated by the buy-sell cycles
\cite{yeung2006}.
The anti-arbitrageurs, 
who sell at a lower price and buy immediately at a higher price,
lose $(1-2\beta)|A_0|^\gamma$ per cycle.
Since the virtual wealths of their strategies are much lower, 
they are outnumbered by the arbitrageurs.
Other agents adopt strategies which have an unequal number
of buy and sell decisions in a period-2 cycle, 
and their maximum or minimum positions prevent
them from participating in the market. 

The behaviors of the arbitrageurs are reminiscent
of those of the privileged traders in real financial markets. 
They place their buying (selling) bids in the market,
stimulating price rises (drops). 
The regime of low market impact corresponds
to the dominance of these traders,
who have the privilege to have their bids cleared
earlier than the other traders, 
and hence enjoy a more favorable price before the price moves significantly.
At the next step, 
they place an opposite bid in the market,
stimulating a price change in the reverse direction,
and recovering their cash (stocks) with a profit margin.

As shown in Fig.~\ref{allWealth}(a), 
the average wealth of the agents increases linearly with time. 
When $\beta=0.5$, 
the agents buy and sell at the same price in the cycle,
resulting in zero change their average wealth
after the transients have subsided,
as shown in Fig.~\ref{allWealth}(d).

\begin{figure*}
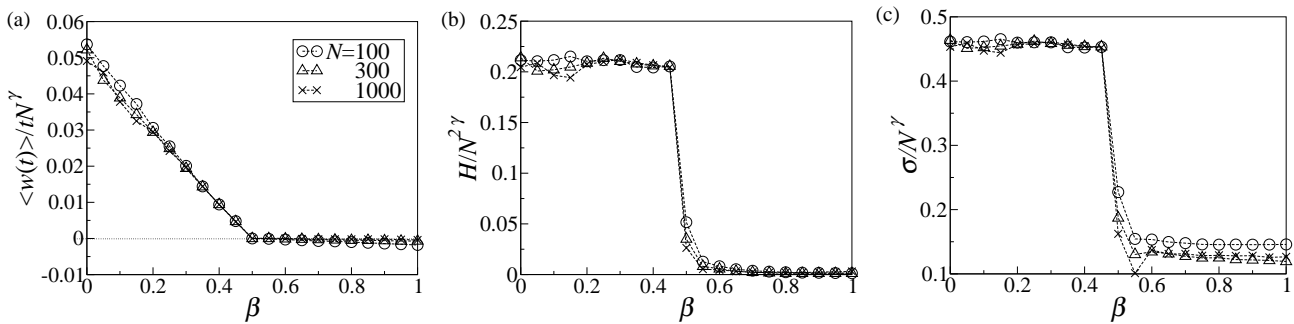

\includegraphics[width=0.31\linewidth]{./wealthVsBeta.eps}~
\includegraphics[width=0.31\linewidth]{./HVsBeta.eps}~
\includegraphics[width=0.31\linewidth]{./sigmaVsBeta.eps}~
\caption{
The dependence on $\beta$ of
(a) the average wealth gain per step,
(b) the predictability, and
(c) the volatility,
for $\gamma=0.5$ and $K=1$. 
Simulations are performed with $N=100$, 300, 1000, $10^5$ steps, and 1000 samples.
}
\label{predictabilityGra}
\end{figure*}

A phase transition takes place at $\beta=0.5$,
since in the regime of large market impact with $\beta>0.5$, 
agents engaged in similar buy and sell cycles are eventually losing
as they are buying at a higher price and selling at a lower price. 
As shown in Fig.~\ref{predictabilityGra}(a), 
the average wealth gain of the agents changes from positive to negative
at $\beta=0.5$ for $\gamma=0.5$ and $K=1$.
The change is accompanied by a change in the slope,
indicating that a continuous phase transition is taking place,
and can be attributed to the disappearance of the arbitrageurs' attractor 
at $\beta=0.5$.

Furthermore, we monitor the predictability $H$ \cite{challet1999}
and the volatility $\sigma$ \cite{savit1999}. 
The former is defined as 
\begin{eqnarray}
        H = \sum_\mu \rho(\mu)\langle \delta P|\mu \rangle^2
\end{eqnarray}
where $\langle \delta P|\mu \rangle$
is the average of the price change conditional on state $\mu$,
and $\rho(\mu)$ is the probability of occurrence of state $\mu$.
The volatility $\sigma$ of price changes is defined as 
\begin{eqnarray}
	\sigma = \sqrt{\langle \delta P^2\rangle-\langle \delta P\rangle^2}.
\end{eqnarray}
Since the dynamics of the arbitrageurs' attractor
is embedded in a low dimensional space,
we expect that the excess demand $A$ scales as $N$ in this regime
\cite{wong2004, wong2005}.
Consequently, 
we expect that $H$ and $\sigma$ scales
as $N^{2\gamma}$ and $N^\gamma$ respectively.
As shown in Figs.~\ref{predictabilityGra}(b)-(c), 
the predictability vanishes and the volatility is greatly reduced
for $\beta$ above 0.5,
supporting the pictures of a phase transition.

\subsection{The Trendsetters' Attractor}

\begin{figure}
\centerline{\epsfig{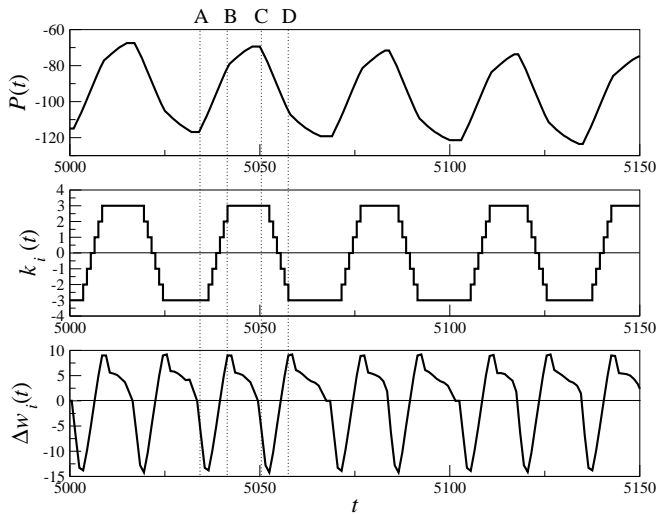}}
\caption{An example of the trendsetters' attractor showing
(a) the price
(b) the position of a typical trendsetter, and
(c) the wealth of a typical trendsetter.
The first to fourth stages start at points A, B, C, D respectively.
Parameters are $m=3$, $s=2$, $K=3$, $\gamma=0.3$, $\beta=0.8$ and $N=1000$.
}
\label{trendSetterGra}
\end{figure} 

\subsubsection{Dynamics}

The phase space is dominated by the {\it trendsetters' attractor}
for sufficiently low $\gamma$.
It is quasiperiodic, 
an example being shown in Fig.~\ref{trendSetterGra}. 
A trendsetter strategy of memory size $m$ consists of
a buying decision responding to a signal
with $m$ consecutive instants of rising price,
and a selling decision responding to a signal
with $m$ consecutive instants of dropping price.
The trendsetters' attractor phase is increasingly prominent 
when the maximum position $K$ increases. 
For a large $K$, 
the agents are allowed to hold multiple short or long positions in hand. 
This increased freedom allows them to gain
by holding a long (short) position
when the price is rising (dropping). 
As shown in Fig.~\ref{trendSetterGra}(a), 
this strategy enables one to gain at all time steps
in a market dominated by the trendsetters' attractor,
except near the points where a rising trend switches to a dropping one, 
or vice versa. 
Consequently, the trendsetter strategy is the most successful strategy
in a trendsetters' attractor,
and the agents holding at least one trendsetter strategy
will make decisions accordingly.
These agents are referred to as the trendsetters.

The dynamics of the attractor consists of four stages:

(1) In the first stage, 
the trendsetters make buying decisions collectively. 
Starting from short positions, 
they accumulate stocks step by step
and switch from minimum to maximum positions in $2K$ steps.
At the end of this stage, 
the positions of the trendsetters reach the maximum, 
and they are restrained from further buying actions.

(2) In the second stage, 
another group of agents takes up the role of collective buying
and pushes the price further up.
This group consists of the {\it fickle} agents,
who fickle their decisions between two strategies.
Through the adaptive process of trading, 
the virtual wealths of the two strategies adjust to such levels
that the selling decisions of a strategy are favoured in the first stage, 
and the buying decisions of another strategy are adopted in the second stage. 
This causes the fickle agents to take buying actions in the second stage, 
pushing the price further up. 
At the end of this stage, 
the positions of the fickle agents reach the maximum, 
and they are restrained from further buying actions. 
The market will then remain quiet with no price movement, 
waiting for a random signal to drive the next stage. 

(3) The third stage is triggered by a randomly generated selling signal, 
and the trendsetters respond by collective selling decisions. 
Starting from long positions, 
they cash in stocks step by step and switch from maximum to minimum positions in $2K$ steps. 
At the end of this stage, 
the positions of the trendsetters reach the minimum, 
and they are restrained from further selling actions.

(4) In the fourth stage, 
the fickle agents switch their strategies to selling, 
since the falling price in the third stage reduces 
the virtual wealth of the buying strategies
and boosts that of the selling strategies.
At the end of this stage, 
the positions of the fickle agents reach the minimum, 
and they are restrained from further selling actions. 
The market will then remain quiet with no price movement, 
waiting for a random signal to drive another cycle.

\subsubsection{Stability}

The stability of the attractor depends on whether 
the trendsetters are able to increase their wealth in a cycle.
Consider the case of $m=2$, for example.
The {\it fast} trendsetters respond to the signals of drop-rise and rise-drop
with buying and selling decisions respectively.
They detect the change in the trend at the earliest possible instant
and become the first group of trendsetters
to benefit in the first and third stages.

On the other hand, 
the slowest trendsetters respond to the signals of drop-rise and rise-drop
with selling and buying decisions respectively. 
They only join the other trendsetters
$m-1$ steps after the onset of a trend.
Hence they lose wealth due to their failure to respond timely 
to the changing trend in the first and third stages. 
 
However, 
since the fickle agents push the price further up and down 
in the second and fourth stages respectively, 
the slowest trendsetters have the opportunity
to regain their wealth through holding long and short positions respectively.
So in other words, if the price change caused by the fickle agents
is sufficient to ensure the slowest trendsetters to gain wealth in a cycle, 
the trendsetters' attractor will become stable.

The stability is achievable
when the price sensitivity $\gamma$ is sufficiently low.
Low values of $\gamma$ imply
that the return in a step is rather insensitive to the excess demand.
Since the virtual wealth of the strategies
held by the fickle agents is lower than that of the trendsetters, 
the population size of the fickle agents
is less than that of the trendsetters.
Hence if the return is insensitive to the excess demand, 
the price change caused by the fickle agents in the second and fourth stages 
will not be too much less
than those caused by the trendsetters in the first and third stages.
The slowest trendsetters will then be able to regain their wealth. 
Consequently, the trendsetters¡¦ attractor is only stable
for sufficiently low values of $\gamma$.

While the analysis for the general case is not available so far,
we have considered  the simple case of $m=2$
which capture the essentials of the agent dynamics.
In this case, 
the number of kinds of agents is few enough for a systematic analysis.
By carefully tracing the decisions of the different types of agents
in this essential model,
one can deduce the price change
and the wealth gained by the different types of agents in the attractor,
and hence the stability condition of the trendsettes' attractor. 
Since the analysis is lengthy,
details will not be presented here.
Nevertheless, we mention that the resultant condition, 
in the form $\beta\leq\beta_{\rm trendsetter}(\gamma)$,
has a remarkable agreement with the simulation results.

It is interesting to consider the stability of the trendsetter attractor in the 
grand-canonical version of our model market,
since it depends on the fickel agents 
who continue to lose wealth to the trendsetters.
Consider the grand-canonical market in which each agent has an extra 0-strategy,
which receives a payoff of $\epsilon$ at every time step and
whose decisions are 0 irrespective of the environment;
$\epsilon$ is the interest rate.
This allows the agents to refrain from playing, 
if their other strategies do not grant them a positive return.
Since the fickle agents in the trendsetter attractor lose wealth,
one might expect that they will withdraws from the market,
destabilizing the attractor.

Surprisingly, 
we find that the trendsetter attractor continues to exist in the 
grand-canonical market. 
We monitor the strategies of the fickle agents,
and find that their strategies are in fact winning strategies 
averaged over time.
Hence,
the fickle agents do not use the 0-strategy.
However,
they switch between the good strategies in an untimely manner,
causing them to lose wealth even though their strategies
are gainning when averaged over time.

Furthermore,
we note that the market behavior is rather sensitive to the interest rate
in the original version of the Minority Game.
For example,
the predictability drops significantly when the interest rate changes
from 0 to a small positive value \cite{challet2001a}.
In our model,
the market behavior in the grand-canonical setting 
is effectively the same as that in the canonical one when the
interest rate is around 0, 
as indicated by the wealth gain per step, 
predictability and volatility. 
Even at an interest rate as high as 0.1, 
the trendsetter attractors continue to exist,
although the wealth gain per step,
predictability and volatility of the market are gradually reduced 
in the negative-sum regime.
The insensitivity to the interest rate in our model is probably due 
to the use of wealth-based payoff schemes,
and the trendsetters' phase is in the regime of positive sum.

\subsubsection{Dependence on $\gamma$}

\begin{figure*}
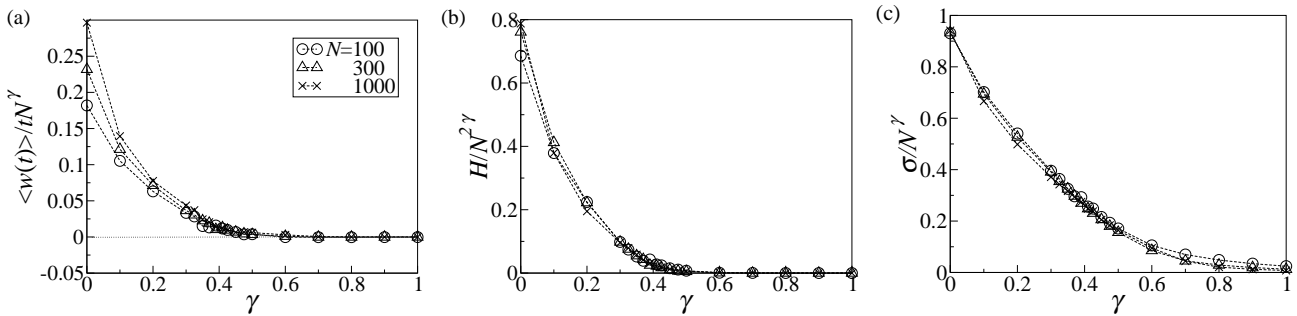

\includegraphics[width=0.31\linewidth]{./wealthVsAlpha.eps}~
\includegraphics[width=0.31\linewidth]{./HVsAlpha.eps}~
\includegraphics[width=0.31\linewidth]{./sigmaVsAlpha.eps}~
\caption{
The dependence on $\gamma$ of
(a) the average wealth gain per step,
(b) the predictability, and
(c) the volatility,
for $\beta=0.8$ and $K=3$. 
Simulations are performed with $N=100$, 300, 1000, $10^6$ steps, and 100 samples.
}
\label{fig:trend-irreg}
\end{figure*}

Figure~\ref{fig:trend-irreg} shows the average wealth gain per step,
the predictability, and the volatility
when $\gamma$ increases across the phase boundary
separating the phases of the trendsetters' and irregular attractors.
There appears a change in the slope of the wealth gain per step
in Fig.~\ref{fig:trend-irreg}(a)
when the phase boundary is crossed,
but no discontinuities seem to appear in the predictability and volatility
in Figs.~\ref{fig:trend-irreg}(b) and (c).
This indicates that the phase transition of the trendsetters' attractor
is milder than that of the arbitrageurs' attractors.

Another interesting feature of the trendsetters' attractor is 
that the price does not return to the same value after one period.
This contributes to a constant growth or decay rate after averaging
over many periods. 
We have also studied the sample-averaged growth
rate of the price,
and found that it also scales as $N^\gamma$.
After rescaling by $N^\gamma$, 
the dependence of the growth rate of the price is qualitatively similar
to those in Fig.~\ref{fig:trend-irreg}(a)-(c).

In the region
where both the trendsetters' and the arbitrageurs' attractors are stable,
we have a {\it mixture phase}. 
The type of attractor reached in the steady state depends on the sample. 
In this mixture phase, 
some trendsetters' attractors are quasistable in the transients, 
and eventually settle as arbitrageurs' attractors in the steady state.

\subsubsection{Relation to other models}

The behavior of the trendsetters' attractor is very similar to the 
bubbles and crashes found in \cite{giardina2003}.
There,
 the trendsetters first create a rising trend,
which subsequently flattens because of their limited buying power.
Then,
the so-called fundamentalists act contrarily to the main trend,
triggering the crash.
It is impiring to note the universality of behavior in these two models,
despite the different ways of introducing the relevant parameters.
In our model, 
the trendsetters' attractor exist only when the price sensitivity $\gamma$
is sufficiently low,
so that the trendsetters and their followers are able to gain
wealth at the end of the rising (dropping) trend by having
long (short) positions.
On the other hand,
the crashes and bubbles in \cite{giardina2003}
are sustainable only when the fraction of investment 
per agent per step is sufficiently small.
We note that both the price sensitivity in our model and the 
investment fraction in \cite{giardina2003} play the role of 
transducing the excess demand to price movement.

Moreover, 
both models contain mechanisms for trend reversal.
In \cite{giardina2003}, 
trend reversals are triggered by the fundamentalists,
whereas in our model trend reversals at the end of rising or dropping
trends become unavoidable since the positions of the 
agents have reached their maximum or minimum.

On the other hand,
our model differs from that of \cite{giardina2003} in having 
fewer parameters,
having the market-makers, 
and having no reference return.
The robustness of the trendsetters' attractor despite these differences
shows that it is ubiquitous phenomenon in certain parameter ranges.
As a further step, 
we have described the role played by the different types of agents 
in the self-organization of these attractors.

The behaviors of the agents in the trendsetters' attractor
resemble those in real financial markets.
The trendsetters respond to a bullish signal
and take buying actions collectively,
creating a bullish market. 
They stop buying due to their finite capital or assessment of risks, 
followed by the fickle agents who try to catch up. 
Eventually the market becomes quiet when all agents 
have exhausted their capital or tolerance to risks. 
Then a bearish signal appears and the trendsetters sell their stocks, 
followed by the fickle agents.

\subsection{The irregular phase}

In the phase diagram of Fig.~\ref{attractorPhaseDia}, 
there is an irregular phase
in which the periodic and quasiperiodic attractors cannot persist.
The trendsetters' attractor appears only intermittently, 
since they are not stable in the long run. 
Other intermittent states are observed. 
Abrupt and sudden changes are found connecting the intermittent states.

\begin{figure}
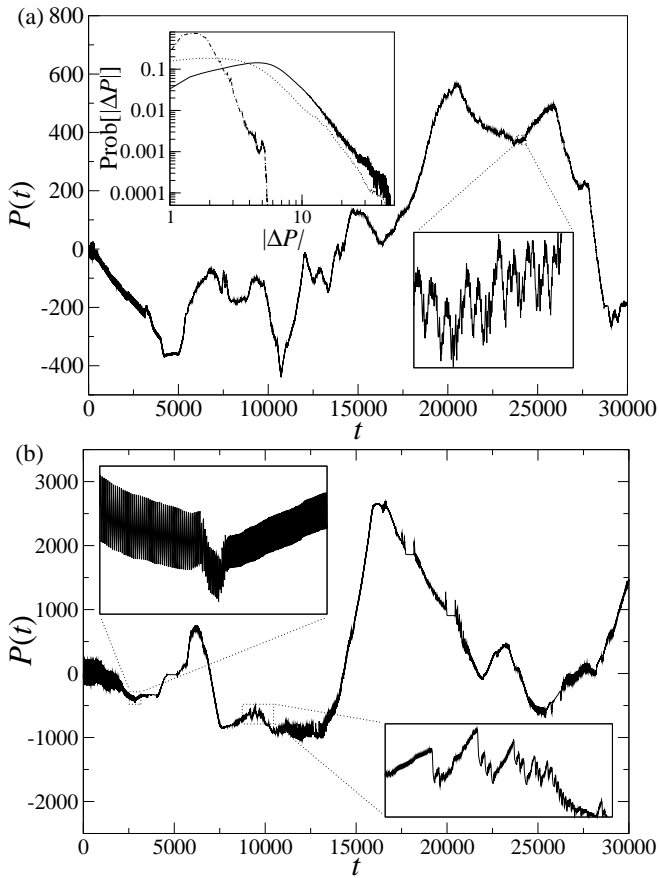

\centerline{\epsfig{figure=priceK1.eps, width=\linewidth}}
\centerline{\epsfig{figure=price.eps, width=\linewidth}}
\caption{(a) An example of the time dependence of the price
at $\beta=0.8$, $m=3$, $s=2$,$\gamma=0.7$, $K=1$ and $N=1000$.
Inset:
The volatility distribution of price
with $N=1000$, $\beta=0.8$ and $m=3$, $(K, \gamma)=(1, 0.3)$
(dash-dotted line),
and $(K, \gamma)=(3, 0.7)$
(dotted line).
Solid line: $N=10000$, $m=6$, $s=2$, $\gamma=0.5$ and $K=3$.
(b) Another example of the time dependence of price at $K=3$, 
other parameters the same as in (a).
}
\label{priceExample}
\end{figure} 

For $\gamma<1$, 
long term price fluctuations embody fluctuations in shorter time scales. 
An example is shown in Fig.~\ref{priceExample}(a). 
The trendsetters' attractors appear
in some transient and intermittent states,
especially in the early stage, 
but not persistently. 
The bottom inset shows the time interval in which the system
does not lie in the trendsetters' or the arbitrageurs' attractors,
showing the fluctuations in short time scales
embodied in longer time scale fluctuations.

The case of $\gamma=1$ is special. 
Since the price change is linearly related
to the global sum of actions $A(t)$,
the price trend in the long run is flat,
consequential to the limit
$\lim_{t'\rightarrow\inf}\sum_{t'}^t A(t') \approx 0$
for finite maximum positions. 
Thus, 
the condition $\gamma<1$ is essential
in reproducing the non-stationary price trend
commonly found in real financial markets.

The inset of Fig.~\ref{priceExample}(a)
shows the corresponding distribution of price change,
with the fat tail of the distribution obeying a power law. 
For the leftmost curve, 
the exponent is approximately -7.4, 
resembling the fat tail volatility distribution
in real financial markets \cite{stanley1999}.
The sharp drop at the end of the tail
is due to finite size limitations in simulations.
The other curve in the same figure shows that increasing $\gamma$ 
causes the distribution of fluctuations to spread out,
and the fat tail decays more gently.

Fig.~\ref{priceExample}(b) shows another example of the time series 
of price in the irregular phase with $K=3$. 
As shown in the top inset of Fig.~\ref{priceExample}(b), 
the system switches from one quasi-periodic cycle to another, 
with an intermediate state in between. 
The trendsetters' attractors appear in the transients, but not persistently. 
The bottom inset of Fig.~\ref{priceExample}(b)
shows that fluctuations are present in multiple time scales.
The price volatility distribution in this region 
confirms a fat tail distribution obeying a power law of exponent $-3.8$,
as shown by the solid line in the inset of Fig.~\ref{priceExample}(a).

\subsection{Positive- and negative-sumness}

\begin{figure}
\centerline{\epsfig{figure=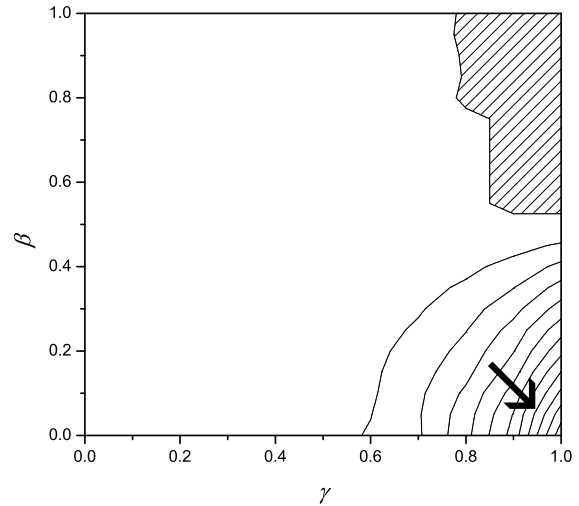, width=\linewidth}}
\caption{
The average wealth gain per step of the agents after equilibration 
as a function of $\gamma$ and $\beta$ for
$m=3$, $s=2$, $K=3$, $N=1000$, $10^6$ steps with 400 samples.
Each coutour represents an increment of 2.
The shaded region corresponds to the region of negative average wealth.
The direction of arrow indicates the the direction of increasing wealth gain.
}
\label{wealthPhaseDia}
\end{figure} 

The average wealth of the agents in the space of $\gamma$ and $\beta$
is shown in Fig.~\ref{wealthPhaseDia}.
For $K=3$,the region of positive average wealth dominates the space.
This means that the market is a {\it positive-sum} game for the agents. 
Much of the positive-sum region
is covered by the region of the two attractors,
as can be seen from a comparison with Fig.~\ref{attractorPhaseDia}. 
Besides that, there exist regions of positive-sums in the irregular phase. 
The behavior of this region resembles the real financial market: 
agents have a positive gain on average, 
while the dynamics is not periodic. 
Furthermore, 
the volatility distribution of the price
follows a fat tail distribution with a power law exponent $-4$
in the case of $K=3$, 
not very far from those observed in real financial markets.

This positive-sumness adds a new perspective to conventional market models, 
such as the original Minority Game, 
which has a negative sum in terms of the number of winners. 
We remark that the positive-sumness is a natural consequence 
of the model where artificial means, 
such as capital injection, 
are not employed to ensure profitability. 
In real markets, 
this is essential to resolving the issue of attracting agents to participate, 
were they given the option.

\begin{figure}
\centerline{\epsfig{figure=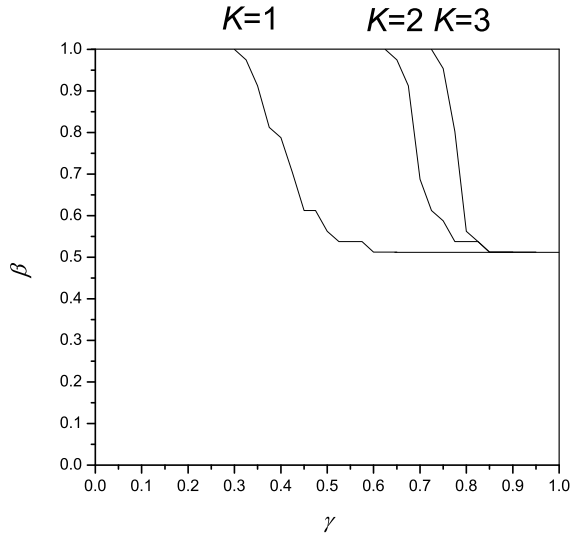, width=\linewidth}}
\caption{
The regions of positive and negative-sumness in the space of $\gamma$ and $\beta$ for
$m=3$, $s=2$. 
Simulations are performed with $N=1000$, $10^5$ steps with 400 samples.
}
\label{wealthPhaseDiaK}
\end{figure} 

As shown in Fig.~\ref{wealthPhaseDia}, 
the average wealth decreases when $\beta$ increases, 
since the market impact reduces the profit of the transactions. 
Furthermore, 
the effect of the market impact on price movements is hardly felt 
if the price sensitivity $\gamma$ is not sufficiently high. 
Hence the region of negative-sumness is only found 
at high values of both $\beta$ and $\gamma$. 
This feature also depends on the maximum position $K$.
In general, the region of negative sum shrinks with $K$,
as shown in Fig.~\ref{wealthPhaseDiaK}.

\subsection{The role of the market-makers}

To understand the origin of the positive-sumness,
we note that supply and demand in the market is not balanced.
When there are more buyers than sellers, 
we implicitly assume someone is willing to provide the extra stocks
for the buyers. 
Likewise,
when there are more sellers than buyers, 
we implicitly assume someone is willing to absorb the extra stocks
from the sellers.
This role of clearing the extra supply and demand at every time step
is played by the market-makers.
At every time step, the market-makers take actions in opposite to
the collective actions of the agents.
It results in an exactly zero-sum game if we sum up the wealth increment
of all the agents and the market-makers.
Although the overall wealth is zero,
the phase diagram in Fig.~\ref{attractorPhaseDia}
actually shows the wealth shifted from the market-maker to the players,
which contribute to the positive-sumness of the players.
It serves as a main allure
for individual players to participate in the market.
However, 
the positive-sumness of the players
implies the negative-sumness of the market-makers,
and the issue of voluntary participation
shifts to why the market-makers have incentives
to participate in the market.
Hence,
there should be additional mechanisms bringing positive gain to both
players and the market-makers.
Before we address this issue in the next section, 
we consider the case of markets without the market-makers.

\subsection{Markets without market-makers}

To model markets without market-makers, 
we balance the supply and demand at every time step
by randomly drawing agents from the majority side
to match the minority side,
whereas the excess demand before matching is still used
in the determination of price change. 
Those who are not fortunate enough to carry out their decisions
are forced to reset their actions to zero 
for the time being.
As a result, 
the number of actual buyers and sellers will be equal at every step.
Among the agents,
the total number of long positions is always equal
to the total number of short positions,
leading to a zero-sum game among the agents.

The price trend in this model is very monotonic.
This can be attributed to the extra buyers (sellers)
whose decisions are rejected by the market.
While their decisions are frustrated,  
their bids nevertheless drive the price up (down), 
leading to a signal of rising (dropping) price at the next step.
With their frustrated decisions, 
they repeat their decisions and buy (sell) again.
The signal of rising (dropping) price continues,
and eventually all agents making the opposite decision are exhausted.
Driven by this extra demand or supply, 
the price goes either up or down continuously, 
but there are no transactions.
This behavior resembles the ``stubborn majority'' in the $\$$-game
\cite{andersen2003}
except that there are no market-makers to be arbitraged in the present case.
The scenario is familiar in the real estates market
of Tokyo in the early nineties or Hong Kong in the late nineties, 
in which real estates prices are driven by wild speculations
to an unaffordable level.
Alternatively, 
a continuously dropping market with no transactions
is similar to the real estates market
after the burst of the bubble.
 
While our model is successful in reproducing these realistic features,
it has not included mechanisms
that terminate the indefinitely rising or dropping trends.
This constitutes a separate issue beyond the scope of this paper.

\section{The Profit of the Market-Makers}
\label{secTC}

In the real market,
the market-makers should have a positive gain
in return for balancing the supply and demand.
This positive gain encourages them to maintain
their service in the market.
It was proposed that the market-makers may lower the risk
of being arbitraged by the players
through adding their inventory to the excess demand
in the determination of price~\cite{andersen2003}.
However,
we found that this method cannot generate profit for the market-maker
in the present model.
Hence, we consider the alternative proposal that the market-makers 
impose a bid-ask spread during their transactions with the players.
That is, 
the transaction prices at time $t$ are $P_{\rm T}(t) \pm S(t)$
for the buying and selling agents respectively,
where $2S(t)$ is the bid-ask spread at time $t$. 
Obviously,
imposing the spread makes it possible
for the market-maker to gain profit from the transactions.
However,
to ensure that both the agents and the market-makers
have incentives to participate,
the bid-ask spread must be determined in such a way
that the wealth of both the agents and the market-makers are stable in time.
Here,
we consider three ways of determining the spread
and compare their effectiveness with respect to this objective.

\begin{figure*}
\includegraphics[width=0.31\linewidth]{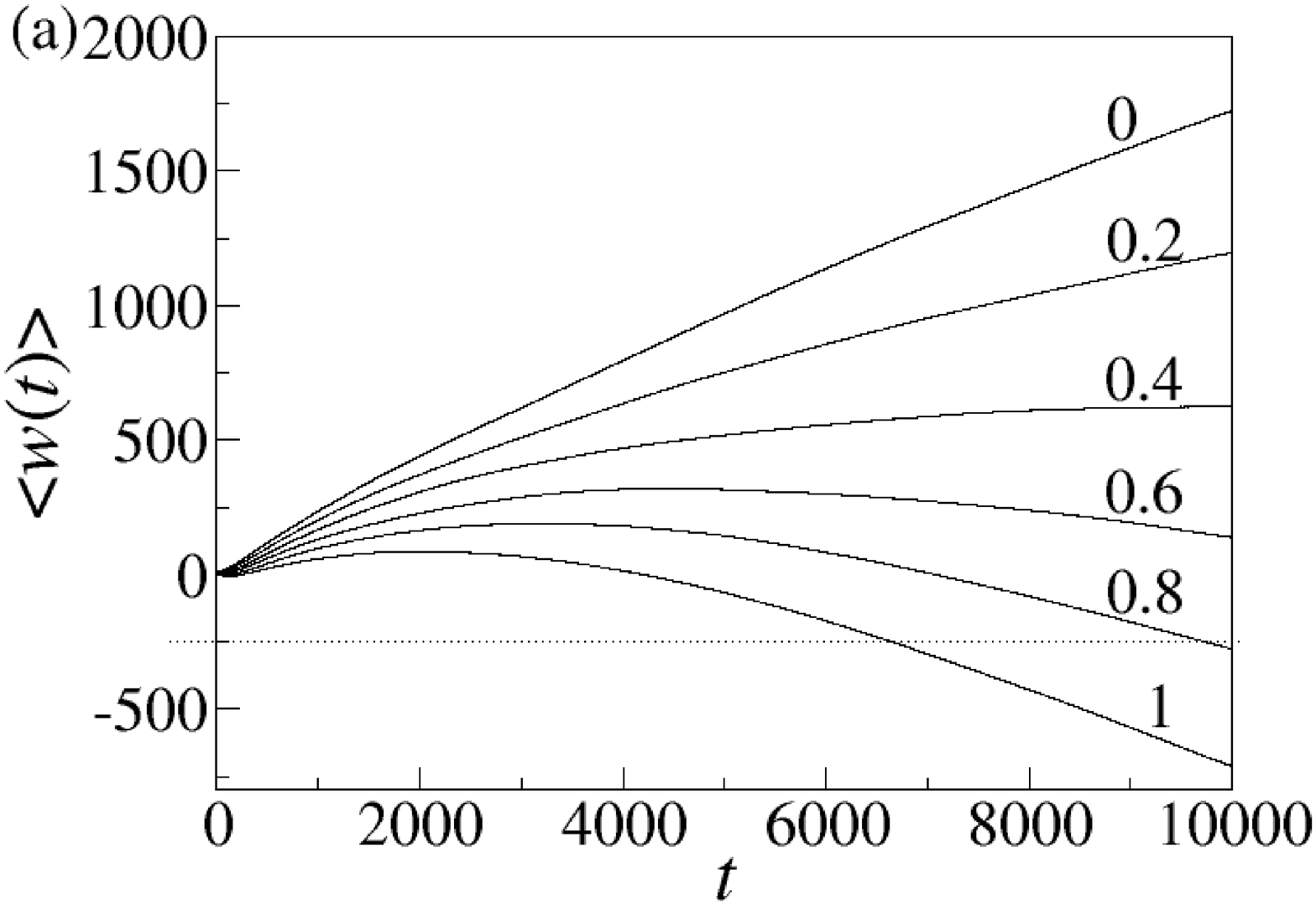}~
\includegraphics[width=0.31\linewidth]{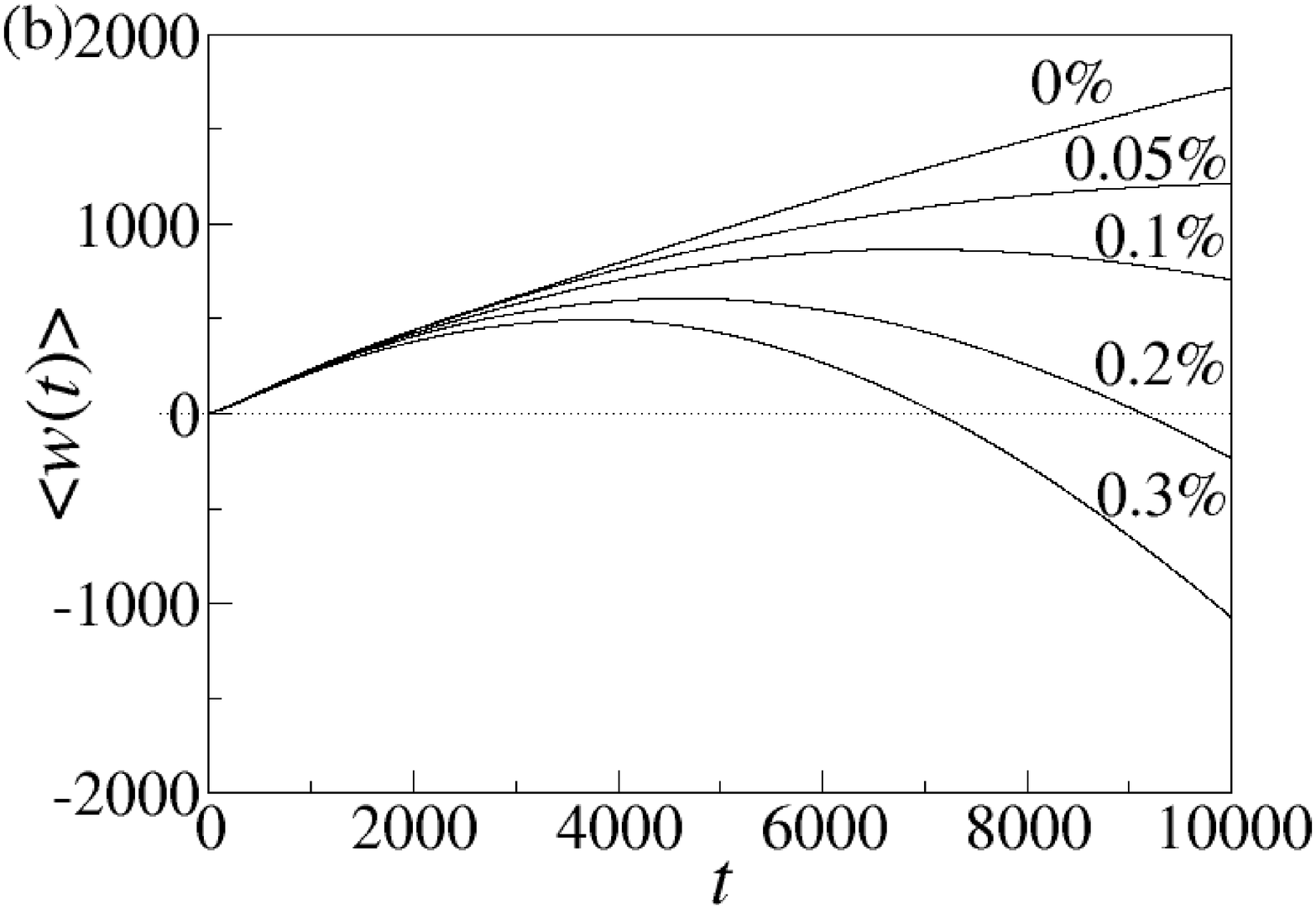}~
\includegraphics[width=0.31\linewidth]{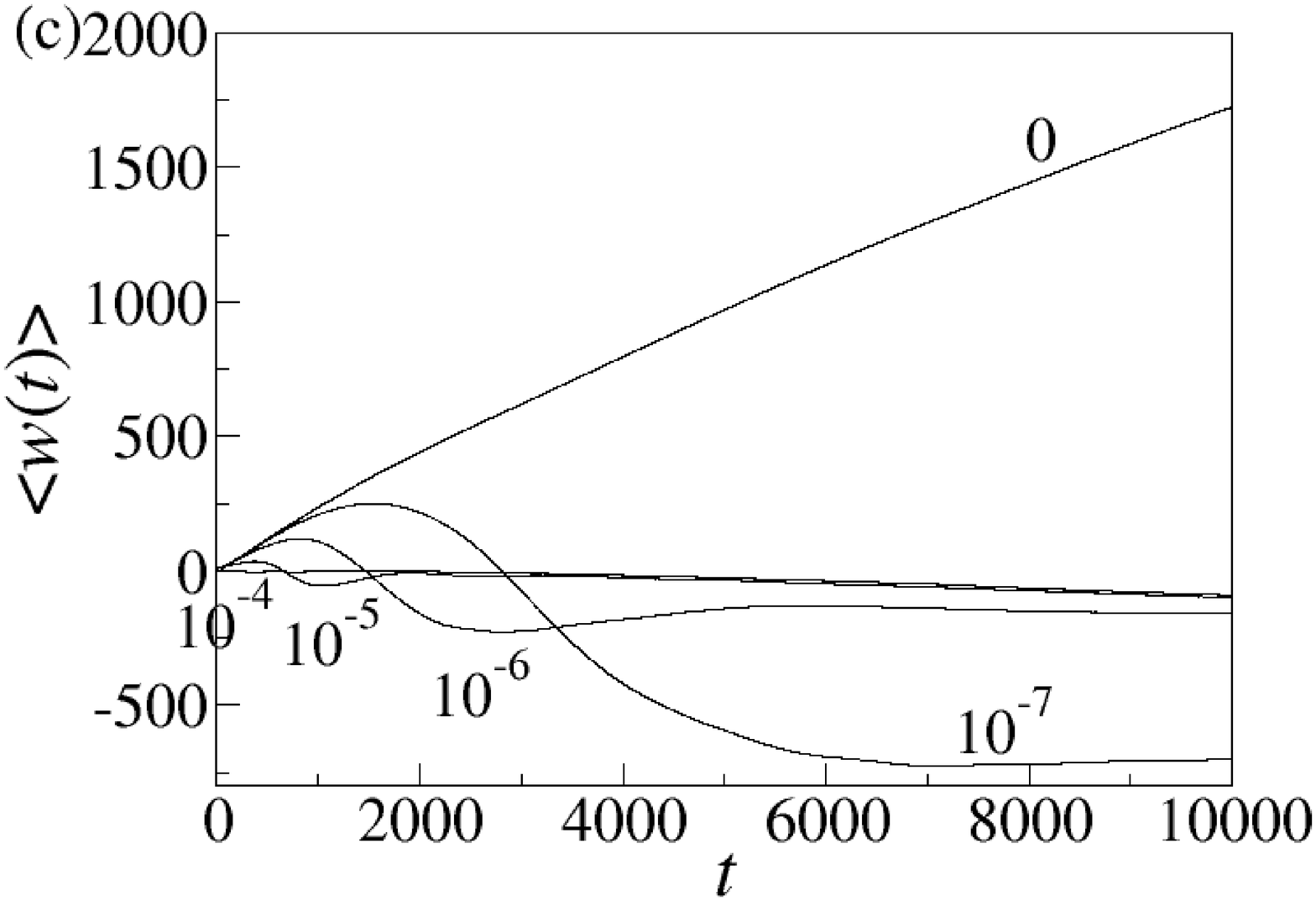}~
\caption{
The time dependence of the average wealth of the agents for $m=3$, $s=2$, 
$\gamma = 0.5$, $\beta = 0.5$, $K=1$, $N=100$ and 1000 samples for
(a) fixed bid-ask spread at the indicated values of $S$,
(b) fixed transaction rate at the indicated values of $R$,
(c) adaptive transaction rate at the indicated values of $\eta$.
}
\label{TCgraph}
\end{figure*} 

1) {\it Fixed bid-ask spread} -
Figure~\ref{TCgraph}(a) shows the average wealth per
agent for different fixed values of $S(t) = S$.
Since the total wealth of the agents and the market-makers sum up to zero,
a positive wealth of the agents
implies a negative wealth of the market-makers,
and vice versa.
We can see that the wealth of the agents decreases with the spread.
When the spread is sufficiently large,
the wealth of the agents become negative at the end of the monitoring period,
implying that the market-makers are getting profit.
However, 
in the framework of fixed transaction costs,
there is only one optimal transaction cost
which yield a zero rate of wealth gain of the agents asymptotically
(such as that around 0.4 in Fig.~\ref{TCgraph}(a)),
and it is difficult for the market to know this optimal cost {\it a priori}.
Even for this optimal cost,
the approach to the asymptotic is too slow.
Consequently,
a fixed bid-ask spread cannot ensure a stable balance of wealth
between the agents and the market-makers.

2) {\it Fixed transaction rate} -
For the market-makers to maintain a profit in the long run, 
we consider a fixed transaction rate $R$ given by $S(t) = R|P_{\rm T}|$.
As shown in Fig.~\ref{TCgraph}(b), 
the rate of change of the average wealth of the agents decreases with time.
This is due to the increasing magnitude of the price with time, 
so that the spread collected by the market-makers increases with time.
Hence, 
for sufficiently high transaction rate
(such as $0.3\%$ in Fig.~\ref{TCgraph}(b)), 
the market-makers gain profit at the end of the monitoring period
shown in Fig.~\ref{TCgraph}(b).
Even for a modest transaction rate
(such as $0.1\%$ in Fig.~\ref{TCgraph}(b)), 
it is anticipated that the agents will eventually lose to the market-makers.
For a low transaction rate (such as $0.05\%$ in Fig.~\ref{TCgraph}(b)), 
we are not certain whether the average wealth of the agents
will continue to increase,
eventually saturate,
or finally decrease beyond the monitoring period.
In any case,
the time it takes to reach a steady balance of wealth between the agents
and the market-makers, 
if it exists at all, 
is apparently prohibitively long.

\begin{figure}
\centerline{\epsfig{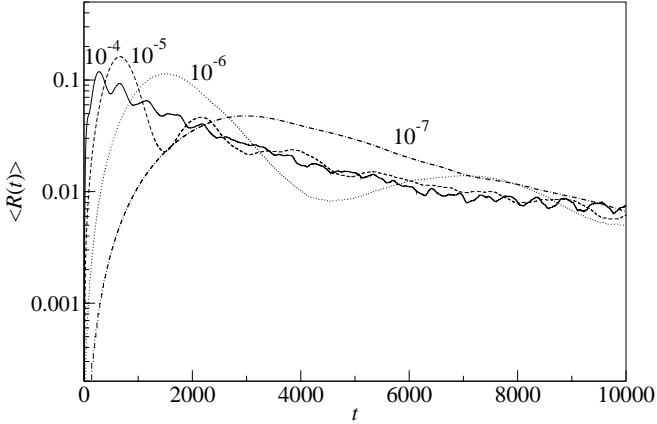}}
\caption{
The time dependence of the sample-averaged transactin rate for
$\eta=10^{-7}, 10^{-6}, 10^{-5}, 10^{-4}$. 
Parameters are the sam as those in Fig.\ref{TCgraph}
}
\label{TRgraph}
\end{figure} 

3) {\it Adaptive transaction rate} -
Since the above two methods cannot ensure
a stable balance of wealth between the agents and the market-makers,
we introduce an adaptive transaction rate as follows.
For a bid-ask spread given by $S(t) = R(t)|P_{\rm T}(t)|$, 
suppose the market-makers decide the transaction rate $R(t)$
so as to attain a total target wealth of $W_{\rm target}(t)$ at time $t$. 
Since their total wealth at time $t$ is $-\sum_i w_i(t)$, 
they would increase (decrease) $R(t)$
if $W_{\rm target}(t)$ is greater (less) than $-\sum_i w_i(t)$. 
Hence, 
we propose an adaptive transaction rate updated according to 
\begin{equation}
        R(t+1) = \max\left\{R(t)+\frac{\eta}{N}\left[W_{\rm target}
        +\sum_i w_i(t)\right], 0\right\},
\end{equation}
where $\eta$ is the learning rate,
dependent on how soon the market-makers
would like to reach their target wealth.
Here, 
we consider the case that the target wealth approaches zero.
This is the scenario in which there is a fierce competition among the 
market-makers.
Results in Fig.~\ref{TCgraph}(c) show that the average wealth of the agents
(and hence that of the market-makers) converges
to a steady value after the transients die out,
the duration of the transient period decreasing with the learning rate.
Figure~\ref{TRgraph} shows the sample-averaged transaction rate, 
indicating that it is necessary to impose a high transaction rate
in the transient period,
before $R(t)$ approaches a universal trend
effectively independent of the learning rate.
This confirms the previous result
that it is not possible to maintain a stable balance of wealth
between the agents and the market-makers using a fixed transaction rate.

\section{New Players in an Evolving Market}
\label{secEvolution}

So far, 
we have found the existence of positive-sumness
in the space of price sensitivity and market impact, 
and that an adaptive transaction rate can ensure
a stable balance of wealth between the agents and the market-makers.
However, 
the model is still a zero-sum game,
and one cannot keep both the agents and the market-makers happy
at the same time.
To ensure participation incentives of both groups, 
we consider in this section {\it open} markets, 
in which underperforming agents leave the market.
They are replaced by new players
who bring their wealth into the market.
Hence, for every replacment of an agent, 
the average growth of wealth of the entire market is equal to the difference
between the average initial wealth of a new player,
minus the average final wealth of an exiting agent.

One worry concerning the issue of participation incentives is that
when the new players enter the market, 
the old players have already adapted their strategies to trade in the market, 
and the new players have disadvantages
in their effort to profit from the market.
If the new players merely play the role of supplying wealth to the market,
without themselves any hope to profit from their participation,
then the new players will not have incentives to participate in the long run.

Here, we consider the following model of an evolving market.
At $t=0$,  
$N$ agents start their trading activities
with their initial wealth set to zero.
At every $T_{\rm ev}$ time steps, 
the agent with the least wealth leave the market.
She is replaced by a new agent with zero initial wealth
and a new set of randomly chosen strategies.

\begin{figure}
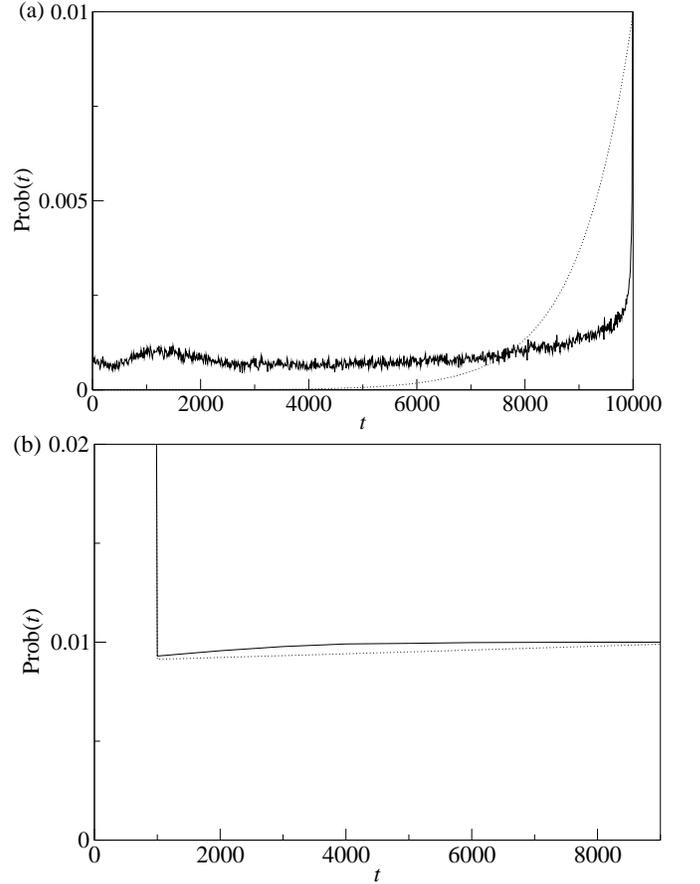

\centerline{\epsfig{figure=tev10_2.eps, width=\linewidth}}
\centerline{\epsfig{figure=tev1000_2.eps, width=\linewidth}}
\caption{
The survival probability at time $T=10000$ of new players entering
the market at time $t$ for 
$m=3$, $s=2$, $K=3$, $\gamma=0.5$, $\beta=0.5$, $W_{\rm target}(t)=0$,
$\eta=10^{-5}$, $N=100$, and 1000 samples, $T_{\rm ev}=$ (a) 10, (b) 1000.
Dotted line: The corresponding survival probability for random evolution
}
\label{survivalProb}
\end{figure} 

\begin{figure}
\centerline{\epsfig{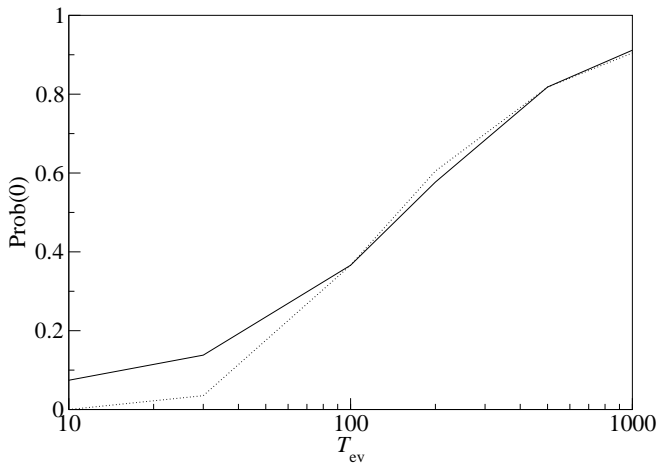}}
\caption{
Solid line: survival probability at time $T=10000$ of the old players
as a function of evolution time $T_{\rm ev}$.
Dotted line: The corresponding survival probability for random evolution.
Parameters are the same as those in Fig.~\ref{survivalProb}.
}
\label{survivalTev}
\end{figure} 

To consider the extent
to which the survival probability of the new players is enhanced, 
we monitor the survival probability at time $T$
of the agents entering the market at time $t$,
and compare it with the survival probability of random evolution,
which is given by $p_{\rm ran}(0) = (1-1/N)^{T/T_{\rm ev}}$ and 
$p_{\rm ran}(t) = (1-1/N)^{(T-t)/T_{\rm ev}}/N$ for $t>0$.
In Fig.~\ref{survivalProb}(a), 
we observe that the agents entering the market
at an early stage of the game
have a higher survival probability than random evolution.
However, 
for fast evolution (low values of $T_{\rm ev}$, 
the most recently entering agents
have a much lower survival probability than random evolution.
Furthermore, 
Fig.~\ref{survivalTev} shows that the survival probability
of the old players is higher
than that of random evolution for low values of $T_{\rm ev}$.
This shows that the enhanced survival probability of the early agents
is not the consequence of gaining wealth from the old players,
but rather taking advantage of the lately entering agents.
This implies that the late corners will have little incentives to 
participate in a fast evolving market.
 
On the other hand,
for a slowly evolving market as shown in Fig.~\ref{survivalProb}(b), 
we see that the survival probability of the new agents
is higher than that of random evolution
for most of the entering time of the new agents.
This shows that the new agents have learned to play as well as, 
or even better than, 
the old players.
 
Figure~\ref{survivalTev} shows the survival probability of the old players
as a function $T_{\rm ev}$.
The survival probability is higher than that of random evolution
for low values of $T_{\rm ev}$,
but is roughly the same for high values of $T_{\rm ev}$.
This again shows that slow evolution provides
a fairer environment for the survival of the new players.

We have also tested the case that the new agents enter the market
with initial wealth equal to the average wealth of the agents
at the moment of entry.
This places the new agents on a much more equal footing
with the old players.
Consequently,
the survival probability of the new agents becomes effectively the 
same as that for random evolution, 
even at the parameters of Fig.~\ref{survivalProb}(a).
This provides another support to the assertion that 
it is possible for the new agents to have participation incentives.

\section{Testing with Hang Seng Index}
\label{secHSI}

To test whether the wealth-based payoff scheme introduced by us
is applicable to real market data,
we replace the self-generated price in the model
by real Hang Seng Index (HSI) data.
For convenience,
we call this wealth-based payoff scheme the Wealth Game.
All the transaction price and history of price change for strategy prediction
are made with respect to the real data.
The data series spans 5045 trading days, 
from 2 January 1987 to 17 May 2007.

\subsection{Tests with Fixed Maxmium Position}

\begin{table}
\begin{tabular}{|l||c|}
\hline
 & Wealth change of strategy $\sigma$ \\
 & after transactions at time $t$ \\
 \hline \hline
 Wealth Game & $k_\sigma(t-1)[P_{\rm T}(t)-P_{\rm T}(t-1)]$ \\
  & $=\sum_{t'}^{t-1}a_\sigma(t')[P_{\rm T}(t)-P_{\rm T}(t-1)]$ \\
  & \\
 Minority Game & $-a_\sigma(t)[P(t+1)-P(t)]$ \\
 & \\
$\$$-Game & $a_\sigma(t-1)[P(t+1)-P(t)]$ \\
 & \\
 Majority Game & $a_\sigma(t)[P(t+1)-P(t)]$ \\
 \hline 
\end{tabular}
\caption{
The payoff change of strategy $\sigma$ after transactions at time $t$ for
the Wealth Game, the Minority Game,
the $\$$-Game and the Majority Game.
Here $a_\sigma(t)$ is the decision made by strategy $\sigma$ at time $t$.
}
\label{tableWealth}
\end{table}

We first consider the case that agents start with zero wealth 
and have fixed maximum positions $K$. 
During this period, 
the agent trades in the market a fictitious stock
whose unit price equals the Hang Seng Index.
She makes her buy, sell and hold decisions using the strategies she holds, 
updating the wealth-based payoffs of her strategies. 
For comparison, 
we also consider the update of payoffs using the Minority Game, 
the $\$$-Game, 
and the Majority Game. 
The payoffs in these other schemes are calculated 
with the excess demand being replaced by the price change $P(t+1)-P(t)$. 
A summary for the schemes is given in Table~\ref{tableWealth}.

\begin{figure}
\centerline{\epsfig{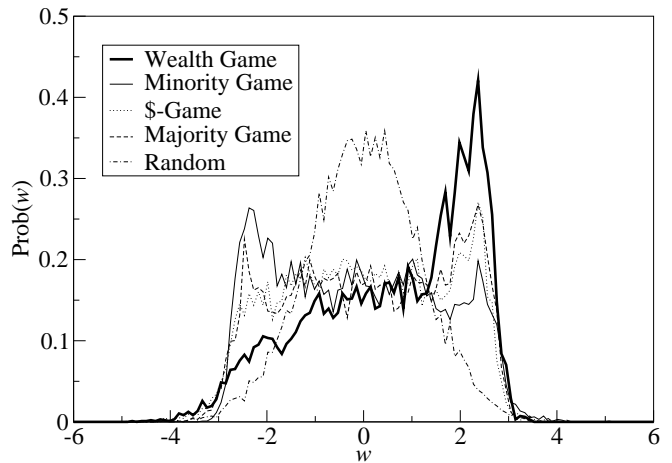}}
\caption{
The final wealth distribution for agents in the Wealth Game,
the Minority Game, the $\$$-Game, and the Majority Game,
in multiples of the HSI closing price of 17 May 2007. 
Parameters: $m=2$, $s=2$, $K=3$, $\beta=0.5$, 10000 agents
for each payoff scheme.
}
\label{wealthDisFixK}
\end{figure} 

Figure~\ref{wealthDisFixK} shows the wealth distribution 
of the four payoff schemes at the end of the trading period. 
The wealth distribution for the Wealth Game
is distinct from the other schemes
in that it has a prominent peak strongly biased
towards the direction of high wealth.
In contrast, 
the Minority Game has a broad peak biased
towards the direction of low wealth.
The average wealth of the payoff schemes are summarized
in Table~\ref{avgBestTable}.
This shows that the Wealth Game produces agents
with the best average performance.

\begin{table}
\begin{tabular}{|l||c|c|c|c|}
\hline
Payoff  schemes& Wealth & Minority & $\$$-Game & Majority \\
& Game & Game& & Game \\
 \hline \hline
 Average wealth & 0.6 & -0.12 & 0.03 & 0.08 \\
 \hline 
 Best wealth & 3.31 & 4.51 & 3.35 & 3.38 \\
 \hline
\end{tabular}
\caption{The average wealth and the wealth of the richest agent in  
the Wealth Game, the Minority Game,
the $\$$-Game and the Majority Game,
in multiples of the HSI closing price of 17 May 2007.}
\label{avgBestTable}
\end{table}

On the other hand, 
Table~\ref{avgBestTable} also shows that the wealth 
of the richest agent produced by the Wealth Game 
is less than those of the other payoff schemes, 
and much less than that of the Minority Game, 
in particular. 
This indicates that while the Wealth Game produces good agents on average, 
it does not encourage the exploration of risky strategies. 
On the other hand, 
the minority-seeking nature of the Minority Game enables the agents 
to explore unconventional strategies giving rise to unusual success, 
but this is achieved through sacrificing 
the performance averaged over the rest of the agents.

\subsection{Tests with Wealth-Based Maximum Position}

To make the model even more realistic, 
we consider the case in which the agents have 
different maximum allowed positions $K_i(t)$
which evolve with time in accordance to their wealth and the current price, 
such that $K_i(t)$ equals the integer part of $\max[w_i(t)/P(t), 0]$. 
The wealth of every agent is initialized
with the same amount of cash wealth $w_i(0)=5P(0)$
to encourage their initial participation in the trading of HSI, 
that is, 
their allowed maximum positions are $K_i(0)=5$ initially.
Compared with the previous case of fixed maximum position, 
one new feature in this test is that as time goes on, 
the wealth of some agents decreases below the current price 
and they are not able to open a new position.

\begin{figure}
\centerline{\epsfig{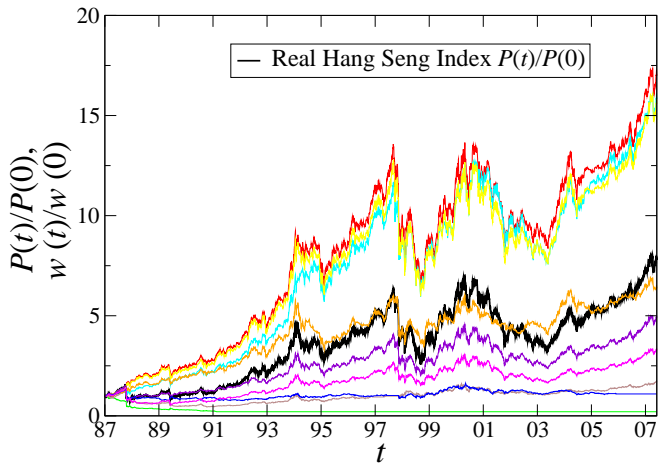}}
\caption{
(Color online) The HSI and the wealth of the three best players, 
five typical players, 
and the worst player in the Wealth Game,
all in multiples of the corresponding values on 2 January 1987.
Parameters: $m=3$, $s=2$, $\beta=0.5$, $N=10000$.
}
\label{HSI}
\end{figure} 

Figure~\ref{HSI} shows the HSI, 
the wealth of the three best players, 
five typical players, 
and the worst player among 10000 players, 
rescaled by their respective initial values.
We make the following observations.
(1) It is possible to have agents
whose wealth grows faster than the price inflation of HSI.
(2) Before the East Asian financial crisis in 1997, 
the best three players adapt to the booming environment
by holding large numbers of stocks (positive positions), 
capitalizing on the rising trend of the price. 
However, 
they are not prepared for the crash. 
Thus, 
they all suffer a great loss in the crash
as they are still holding high long positions immediately before the crash.
(3) For the typical players, 
many of them are already lagging behind the price inflation, 
and thus are not rich enough to hold large numbers of stocks. 
As a result, 
they suffer less than the best players from the crash.
(4) Some players have their wealth eventually falling below the price. 
This limits them from opening new positions. 
Their wealth becomes frozen.

\begin{figure}
\centerline{\epsfig{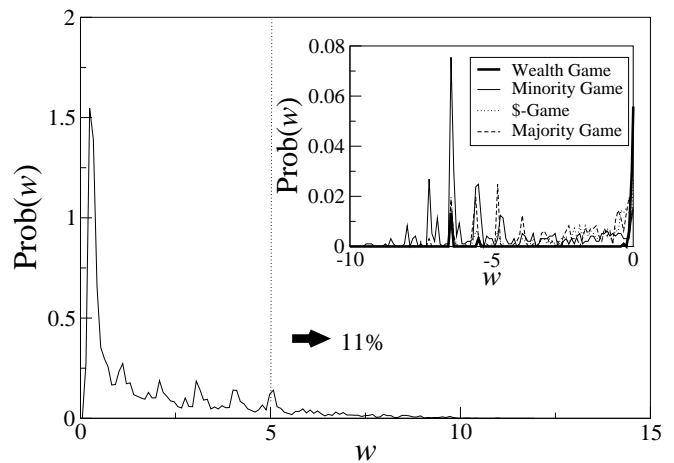}}
\caption{
The distribution of wealth, 
in multiples of the HSI closing price of 17 May 2007, 
among the 10000 players in the Wealth Game with $m=3$, $s=2$, $\beta=0.5$.
Inset: 
The distribution of negative wealth, 
in multiples of the HSI closing price of 17 May 2007, 
among the 10000 players in the Wealth Game, 
the Minority Game, 
the $\$$-Game, 
and the Majority Game,
with $m=3$, $s=2$, $\beta=0.5$. 
The restriction of having at least one buy and one sell decision per strategy
is relaxed.
}
\label{wealthDisVariableK}
\end{figure} 

Figure~\ref{wealthDisVariableK} shows the distribution
of final wealth of the agents
after trading over 20 years of HSI in this experiment.
About $10\%$ of the agents are {\it gaining agents}, 
that is, 
with their wealth growing faster than the price inflation of HSI, 
as indicated by the tail of the distribution
to the right of the vertical line in Fig.~\ref{wealthDisVariableK}.
The wealth of other agents grows slower than the HSI. 
Peaks near integer values are observed, 
corresponding to agents whose wealth 
are clustered around their maximum allowed positions.

\begin{table}
\begin{tabular}{|l||c|c|c|c|}
\hline
Payoff  schemes& Wealth & Minority & $\$$-Game & Majority \\
& Game & Game& & Game \\
 \hline \hline
 Average wealth & 1.85 & 1.47 & 1.40 & 1.31 \\
 \hline 
 $\%$ gaining agents & 10.4 & 8.68 & 6.40 & 5.45 \\
 \hline
 Worst wealth & 0.13 & 0.13 & 0.13 & 0.12 \\
 \hline
 $\%$ bankruptcy & 0 & 0 & 0 & 0 \\
 \hline
 Best wealth & 11.16 & 17.15 & 11.13 & 10.88 \\
 \hline
\end{tabular}
\caption{The performance of the agents in   
the Wealth Game, the Minority Game,
the $\$$-Game and the Majority Game, 
in multiples of the HSI closing price of 17 May 2007.}
\label{avgGainTable}
\end{table}

The Wealth Game is compared with other conventional payoff schemes. 
A control experiment is also done by agents randomly making decisions. 
Table~\ref{avgGainTable} shows that among the four payoff schemes, 
the Wealth Game has the best average performance, 
as evident from the average wealth and the fraction of gaining agents. 
As a benchmark, 
there are $0.03\%$ of gaining agents in the case of random choices.

Another advantage of the Wealth Game can be seen 
by relaxing the constraint that the strategies held by the agents must have 
at least one buy and one sell decision. 
This constraint prevents an agent from getting trapped
in a long (short) position in a bearish (bullish) market,
and hence greatly reduces bankruptcy, that is, negative wealth, 
as illustrated in Table~\ref{avgGainTable} by the worst wealth being positive 
and the bankruptcy rate being 0.

However, 
removing this constraint significantly increases the bankruptcy rate. 
This is especially significant for the Minority Game, 
where the bankruptcy rate is highest among the four payoff schemes, 
and the poorest agents have largely negative wealth,
as shown in the inset of Fig~\ref{wealthDisVariableK}.
On the other hand, 
the same figure shows that the bankruptcy rate for the Wealth Game
is much lower,
and the poorest agents only have a marginally negative wealth. 
Indeed, 
the bankruptcy rates are $0.7\%, 4.7\%, 2.6\%$, and $3.0\%$ for the Wealth, 
Minority, $\$$-, and Majority Games respectively. 
This advantage of the significantly reduced bankruptcy rate 
is probably due to the faithfulness of the Wealth Game payoff
in reflecting the wealth of an agent.

On the other hand, 
the Minority Game is able to produce a small number of agents
with best performances,
as shown in Table~\ref{avgGainTable}. 
This indicates that the Minority Game rewards more risky strategies 
that can result in a small number of successful strategies, 
a conclusion consistent with the results
in the case of fixed maximum position.

At first sight,
it does not appear surprising that the agents in the Wealth Game perform
better than other players,
since their wealth is updated with the rule of this model. 
However, 
the following two tests provide further insights about the 
behaviors of the different payoff schemes.

First, 
we test the payoff schemes on the HSI data for a trendy period.
We consider the period of recovery from SARS, 
starting from the bottom on Apirl 25, 2003 to May 17, 2007.
The price trend is rather monotonic,
making it easy for strategies to adopt.
In this case, 
the Wealth Game has the best average wealth,
and the Minority Game the worst.

Next, 
we consider a rugged period from July 15, 1996 to March 21, 2001
(one year before and after the peaks at July 15, 1997 and 
March 21, 2000 respectively), 
covering both the East Asian financial crisis and the dot-com bubble.
During this period, 
the general trend of the HSI consists of two rises and two falls,
making it rather difficult for the strategies to adopt. 
In this case, the Minority Game performs much better than the other 
payoff schemes,
and the Wealth Game has the poorest average wealth.

In summary,
the Wealth Game performs better than the Minority Game 
in trendy periods of HSI, 
but worse in rugged periods.
This is probably due to the use of positions $k_\xi(t+1)$ 
rather than the action $a_\xi(t)$ in the calculation of the payoff of strategy $\xi$.
This gives a stronger history dependence of the Wealth Game,
which favors its perfomance in trendy markets.
However, this is probably the same reason that makes it face more difficulties 
in periods with many trend reversals.
Overall,
in a time scale as long as 20 years,
the price time series of most financial markets have rising trends such as the one 
in HSI, 
so that payoff schemes with strong history dependence still have overall advantages.

\section{Conclusion}
\label{secConclusion}

We have considered a family of models of the financial market
in which all parties involved,
namely, 
the agents, 
the market-makers, 
and the new players,
find incentives to participate.
This represents a departure from the conventional Minority Games,
whose focus does not reflect the positive-sumness of financial markets.
The pre-requisite of this departure
is the direct usage of wealth as both the payoff
for evaluating strategies and a measure of the success of individual agents,
the latter allowing for the possibility of positive-sumness of the market.

The wealth-based payoff scheme is successful
in reproducing collective behaviors of the agents
resembling those in real markets.
Specifically,
the price sensitivity $\gamma$ and the market impact $\beta$
are introduced to fine tune, respectively, 
the transduction from the excess demand to the price dynamics, 
and the reaction of collective decisions on individuals.
In the phase space of $\gamma$ and $\beta$, 
several behaviors resembling those in real financial markets emerge.
In the arbitrageurs' attractor, 
alternate buy and sell cycles emerge when the market impact is low,
analogous to the arbitraging activities of privileged traders
in real financial markets.

In the trendsetters' attractor,
quasiperiodic cycles of price rises and falls
are created by the trendsetters and followed by the fickle agents.
This coordinated behavior has a direct correspondence
to the roles played by the different kinds of agents in the real market.
It demonstrates the importance of using virtual wealth
to measure the success of strategies.
Since the virtual wealth depends on the history of decisions, 
it embeds a longer memory in strategy selection, 
thus inducing a greater sophistication in the collective behavior
such as that exhibited in the trendsetters' attractor.

The phases of the arbitrageurs' and trendsetters' attractors
are bounded by phase transition lines typical of statistical physics.
Their quasiperiodicity demonstrates
the robustness of the players' coordinated behavior.
On the other hand,
while they reproduce agent behaviors plausible in real markets,
it seems that their rigid periodicities
cannot be reproduced in real markets.
It will be interesting to study
whether dynamical behaviors of the same nature
can be generated by models with more heterogeneity, 
say, in the memory sizes and maximum positions of the agents.

The phase diagram also consists of the irregular phase,
in which the time series of the price
consists of intermittent quasiperiodic states
connected by abrupt changes. 
Simple or linear price trends are not observed, 
making the system dynamically interesting.
Fluctuations  in the long time scales
embody fluctuations in shorter time scales.
This is the regime with price time series resembling real markets most.

The phase diagram is rather robust to variations of the model.
For example, 
in the grand-canonical version in which agents with poor strategies are allowed 
to refrain from playing,
we found that the trendsetters' attractor is still sustained by the presence 
of fickle agents.
We have checked that the behavior of the other phases are 
qualitatively the same in the grand-canonical game.

An interesting question is what parameter range of $\gamma$ and $\beta$
should correspond to the real market.
The attraction of agent-based models is its ability to relate
adjustable microscopic mechanism with macroscopic market behavior,
thereby giving rise to self-organizing mechanisms for market evolution.
Self-organization of markets have been proposed for their states of criticality 
\cite{challet2001b} and efficiency \cite{giardina2003}.
Here, 
we propose a possible mechanism for the market to self-organize to the irregular phase
resembling the real market.

Suppose the market is in the trendsetters' phase with a low value of $\gamma$.
Since the market is positively summed in this phase,
the agents are encouraged to act boldly in bidding and asking
to maximize their expected wealth gain.
This leads to an increase in the price sensitivity.
The trend will continue until the market behavior enters the irregular phase.
Near the line of zero sum in the irregular phase,
the reduced wealth gain of the agents discourages them
from increasing their boldness in their bids and asks,
and the value of $\gamma$ reaches a steady value.
The mechanism is similar to the self-organization of the investment level to an
efficient state in \cite{giardina2003}.

The self-organization of the market impact $\beta$
might be related to the activities of the arbitrageurs.
Suppose the market is in the arbitrageurs' phase with a low value of $\beta$.
The profitable arbitraging opportunities attract more arbitrageurs to enter
the market.
Consequently, 
each individual arbitrageur is no longer as privileged as before,
since their decisions will exert market impact on each other.
This results in an increase in the market impact parameter.
The trend wil continue until the market behavior enters the irregular phase.
In practice, 
individual agents experience different market impact due to their different reaction
times and transaction volumes.
Models with individualized market impact should be studied in the future
to further clarify this issue of self-organization. 

Returning to the issue of participation incentives, 
the phase space has a large region of positive-sumness.
This positive-sumness provides incentives
for the agents to play in the market.
In the absence of new players joining the market, 
their wealth comes from the market-makers
who play the role of balancing the supply and demand of the market.

The participation incentives of the market-makers
come from the bid-ask spread during their transactions with the players.
We found that both the fixed transaction cost and fixed transaction rate 
cannot ensure a stable balance of wealth
between the agents and the market-makers.
Rather,
an adaptive transaction rate avoids this problem.

The picture of paticipation incentives is completed
by considering open markets,
in which underperforming agents are replaced by new agents.
This converts the zero-sum game of the agents and market-makers
to one with positive sum.
The participation incentive of the new players may arise
from their chances to learn the strategies of the more successful players.
We found that these agents
have survival probabilities matching those of random evolution, 
provided that the evolution is sufficiently slow.

Finally, we have tested the wealth-based payoff scheme 
by playing the Wealth Game on HSI data over 20 years, 
and compared its performance with the Minority Game,
the $\$$-Game, and the Majority Game.
We found that the Wealth Game produces agents
with much better average performance than other games,
and significantly reduces the bankruptcy rate
for agents holding poor strategies.
This is probably due to the faithfulness of the payoff scheme 
in reflecting the wealth associated with a strategy.
Further tests on other markets are needed to experiment the 
applicability of wealth-based payoff schemes and 
will be reported elsewhere.

On the other hand, the Minority Game is able to produce
a small number of extremely rich agents,
whose wealth is better than the richest agents
produced by other payoff schemes.
This indicates that while the Wealth Game produces good agents on average, 
it does not reward the exploration of risky strategies, 
whereas the minority-seeking nature of the Minority Game 
enables the agents to explore unconventional strategies
giving rise to unusual success,
but this is achieved through sacrificing the performance
averaged over the rest of the agents.

The strengths and weaknesses of the different payoff schemes
are also revealed in their performances in trendy and rugged periods.
The Wealth Game performs well when history can provide a guide,
but meets difficulties in markets with many trend reversals.

These performance characteristics of the wealth-based payoff scheme 
can be traced to the different ways virtual scores are updated
in the payoff schemes.
Among the many differences between the Wealth Game and the Minority Game, 
the most essential one is that the position $k_\sigma(t-1)$ 
replaces the action $a_\sigma(t)$
in the calculation of the virtual score in Table~\ref{tableWealth}.
In the Wealth Game, agents assess their gains by their positions 
rather than one-step decisions. 
Since the position of a series of actions of a strategy
is equal to its historical sum,
position-dependent assessment schemes embed a longer memory scale 
in the decision making process of agents than those in the Minority Game.

Interesting issues arising from this comparative study remain. 
First, wealth-based payoff schemes are able
to explain the behavior of good investors statistically,
but since the extremely successful investors
are produced by the Minority Game
rather than the Wealth Game, 
it is possible that the behavior of the group of rich agents
and the few extremely rich agents
have to be modeled differently, 
especially with respect to the readiness to take risks. 
Second, it would be interesting to formulate integrated payoff schemes, 
so that the agents can both learn from history during trendy periods, 
and be ready to explore risky but potentially successful strategies
during risky periods.

\section*{Acknowledgements}
We thank David Sherrington and Andrea de Martino for meaningful discussions.
This work is partially supported by the Research Grant Council of Hong Kong
(Grant Nos. HKUST603606 and HKUST603607)


\end{document}